\documentclass[preprint,prb,showpacs,preprintnumbers,amsmath,amssymb,endfloats*]{revtex4}
\usepackage[latin1]{inputenc}
\usepackage{graphicx}                          % Include figure files
\usepackage{dcolumn}                           % Align table columns on decimal point
\usepackage{bm}                                % bold math

\newcommand{\myvec}[1]{\vec{#1}}               % Format fuer Vektorgroessen
\newcommand{\mymatrix}[1]{\underline{#1}}      % Format fuer Matrizen
\newcommand{\cc}{\text{c}}                     % colloids
\newcommand{\pp}{\text{p}}                     % polymers
\newcommand{\crit}{\text{crit}}
\newcommand{\rma}{\text{A}}
\newcommand{\rmb}{\text{B}}
\newcommand{\kb}{k_\text{B}}

\begin{document}

%\preprint{October 2008}

\title{Statics and Dynamics of Colloid-Polymer Mixtures Near Their Critical Point of Phase Separation:\\A Computer Simulation Study of a Continuous AO Model}

\author{Jochen Zausch}
\affiliation{Institut f\"ur Physik, Johannes Gutenberg-Universit\"at Mainz, 55099 Mainz, Staudinger Weg 7, Germany}
\author{Peter Virnau}
\affiliation{Institut f\"ur Physik, Johannes Gutenberg-Universit\"at Mainz, 55099 Mainz, Staudinger Weg 7, Germany}
\author{J\"urgen Horbach}
\affiliation{Institut f\"ur Materialphysik im Weltraum, Deutsches Zentrum f\"ur Luft- und Raumfahrt (DLR), 
51170 K\"oln, Germany}

\author{Richard L.~Vink}
\affiliation{Institut f\"ur theoretische Physik, Georg August-Universit\"at G\"ottingen, Friedrich-Hund-Platz 1, 
37077 G\"ottingen, Germany}

\author{Kurt Binder}
\affiliation{Institut f\"ur Physik, Johannes Gutenberg-Universit\"at Mainz, 55099 Mainz, Staudinger Weg 7, Germany}

\date{\today}

\begin{abstract}
We propose a new coarse-grained model for the description of liquid-vapor
phase separation of colloid-polymer mixtures.  The hard-sphere
repulsion between colloids and between colloids and polymers, which
is used in the well-known Asakura-Oosawa (AO) model, is replaced by
Weeks-Chandler-Anderson potentials.  Similarly, a soft potential of
height comparable to thermal energy is used for the polymer-polymer
interaction, rather than treating polymers as ideal gas particles.
It is shown by grand-canonical Monte Carlo simulations that this model
leads to a coexistence curve that almost coincides with that of the AO
model and the Ising critical behavior of static quantities is reproduced.
Then the main advantage of the model is exploited --- its suitability
for Molecular Dynamics simulations --- to study the dynamics of mean
square displacements of the particles, transport coefficients such as the
self-diffusion and interdiffusion coefficients, and dynamic structure
factors.  While the self-diffusion of polymers increases slightly
when the critical point is approached, the self-diffusion of colloids
decreases and at criticality the colloid self-diffusion coefficient is
about a factor of 10 smaller than that of the polymers.  Critical slowing
down of interdiffusion is observed, which is qualitatively similar to
symmetric binary Lennard-Jones mixtures, for which no dynamic asymmetry
of self-diffusion coefficients occurs.
\end{abstract}
%

%\pacs{}

\maketitle

\section{Introduction}
In the last few decades colloidal dispersions have been studied
intensively as model systems for the structure and phase behavior of
fluids and solids.  The large size of the colloidal particles allows for
additional experimental techniques which are not applicable for atomistic
or molecular systems.  Moreover, the colloid-colloid interactions can
be ``tuned'' to a large extent \cite{r1,r2,r3,r4,r5}.  For example,
individual colloidal particles can be tracked through space in real time
using confocal microscopy \cite{r6}.  In colloid-polymer mixtures, where
the depletion attraction between the colloids caused by the polymers
\cite{r7a,r7b,r8} can lead to a liquid-vapor type phase separation
\cite{r9,r10,r11}, statics and dynamics of capillary wave-type interfacial
fluctuations can be observed in real space \cite{r12}.  Wetting layers
of the walls of containers can be studied in detail \cite{r13,r14,r15},
and critical fluctuations can also be seen directly in optical microscope
observations \cite{r16}.  Very interesting nonequilibrium studies are
also possible, such as shear-induced narrowing of interfacial widths
\cite{r17} and studies of spinodal decomposition \cite{r18}.

In view of this wealth of experimental data on static and dynamic behavior
relating to liquid-vapor type phase separation in colloid-polymer
mixtures, it is also desirable to provide a detailed theoretical
understanding of these phenomena.  In fact, many static aspects
(including the understanding of the phase diagram and bulk critical
behavior \cite{r19a,r19b,r20}, interfacial fluctuations \cite{r21}
and interface localization transitions \cite{r22,r23}, capillary
condensation/evaporation \cite{r23,r24a,r24b,r24c,r25,r26} and wetting
\cite{r27,r28,r29,r30}) can all be understood by the simple Asakura-Oosawa
(AO) \cite{r7a,r7b,r8} model, at least qualitatively.  In this model
colloids and polymers are described as spheres of radius $R_\cc$
and $R_\pp$, respectively.  While there is a hard core interaction
of the colloids both among each other and also with the polymers,
the polymer-polymer interaction is assumed to be strictly zero.  Thus,
a suspension without any colloids but only polymers is just treated as
an ideal gas of point particles which are located at the center of mass
of the polymer coils.

This model is very attractive due to its simplicity.  It allows for
various elegant analytical approximations \cite{r10,r28,r29,r30,r31}
as well as for efficient Monte Carlo simulation techniques
\cite{r19a,r19b,r20,r21,r22,r23,r24a,r24b,r24c,r25,r26,r27}.  However,
the assumption that polymers do not interact with each other at
all makes the AO model unsuitable for studying dynamical aspects of
colloid-polymer mixtures.  Thus, a different model is required to
complement the corresponding very interesting experiments mentioned
above \cite{r12,r16,r17,r18}.

Occasionally, computer simulations have been performed where the
polymers were modeled explicitly as chain molecules either on the
lattice \cite{r32,r33a,r33b} or as bead-spring-type chains in the
continuum \cite{r34}.  In general, these models are restricted to rather
small chain lengths in order to keep the numerical effort manageable.
In addition, only particle sizes in the nanometer range can be treated.
However, one can use these simulations \cite{r33a,r33b} to justify an
effective interaction between two polymer coils.  Thus, polymers are
described as soft particles which can ``sit on top of each other'', but
not without energy cost.  The usefulness of such an effective potential
has been amply demonstrated \cite{r33a,r33b,r35a,r35b,r36}.

This consideration is the motivation for the present study.  We define
a model (Sec.~\ref{sec:model}) which has a soft interaction potential
between polymers, too, and is particularly convenient for both Monte
Carlo \cite{r37,r38} and Molecular Dynamics \cite{r38,r39} simulations.
In Sec.~\ref{sec:statics} the static properties of the model are evaluated
and compared to corresponding results \cite{r19a,r19b} for the standard AO
model \cite{r7a,r7b,r8}. Section~\ref{sec:dynamics} presents our data for
the mean square displacements of the particles as well as intermediate
scattering functions.  We also discuss the resulting self-diffusion and
interdiffusion coefficients while Sec.~\ref{sec:conclusion} summarizes
our conclusions.

\section{A soft variant of the AO model}\label{sec:model}
A potential of type $U(r)=U_0\exp[-(r/R_g)^2]$ describes the effective
interaction between two polymer coils in dilute solution under good
solvent conditions.  This result can be obtained by calculating
the partition function of the two chains under the constraint that
the distance $\myvec{r}$ between the centers of mass of the coils
is fixed.  The prefactor $U_0$ is of the order of the thermal energy
\cite{r33a,r33b,r35a,r35b} and $R_g$ is the radius of gyration of the
chains.  Similarly, the interaction potential between a polymer chain
and a colloidal particle can be obtained.

However, the situation becomes slightly more involved at higher polymer
concentrations where many coils overlap and the temperature of the polymer
solution (in comparison with the Theta temperature \cite{r40}) also
plays a role.  Then it is no longer possible to give a simple explicit
description for the polymer-polymer interaction from first principles.
Additionally, it is more convenient for computer simulations to have
a potential which is strictly zero if $r$ exceeds some cutoff $r_c$.
Therfore, we did not use any of the approximated effective potentials
derived in the analytical work \cite{r33a,r33b,r35a,r35b}.  Instead we
chose a potential that has qualitatively similar properties, but is
optimal for our simulation purposes.  For the colloid-colloid and
colloid-polymer potential we took the Weeks-Chandler-Anderson (WCA)
potential \cite{r41}, modified by a smoothing function $S$
\begin{equation}\label{eq:ljpotential}
U_{\alpha\beta}=
4\epsilon_{\alpha\beta}
\left[\left(\frac{\sigma_{\alpha\beta}}{r}\right)^{12}
-\left(\frac{\sigma_{\alpha\beta}}{r}\right)^6+\frac{1}{4}\right]S\,,
\end{equation}
with
\begin{equation}\label{eq:smoothing}
S=\frac{(r-r_{c,\alpha\beta})^4}{h^4+(r-r_{c,\alpha\beta})^4}\,.
\end{equation}
Here, $\epsilon_{\alpha\beta}$ controls the strength
and $\sigma_{\alpha\beta}$ the range of the (repulsive)
interaction potential which becomes zero at $r_{c,\alpha\beta}$
and stays identically zero for $r\ge r_{c,\alpha\beta}$ with
$r_{c,\alpha\beta}=2^{1/6}\sigma_{\alpha\beta}$.  Following previous
work on the AO model \cite{r19a,r19b,r20,r21,r22,r23,r25,r26}, we chose
a size ratio $q=\sigma_{\pp\pp}/\sigma_{\cc\cc}=0.8$ between polymers
and colloidal particles and
\begin{equation}\label{eq:ljsigma}
\sigma_{\cc\pp}=0.5(\sigma_{\cc\cc}+\sigma_{\pp\pp})=
0.9\sigma_{\cc\cc}\,.
\end{equation}
The parameter $h$ of the smoothing function is taken as
$h=10^{-2}\sigma_{\cc\cc}$ and $\epsilon_{\cc\cc}=\epsilon_{\cc\pp}=1$.
In the following, we choose units such that $\kb T=1$ and
$\sigma_{\cc\cc}=1$.  Note that the smoothing function is needed in
Eq.~\eqref{eq:ljpotential} such that $U_{\alpha\beta}(r)$ becomes twofold
differentiable at $r_{c,\cc\cc}$ and $r_{c,\cc\pp}$ without affecting
the potential significantly for distances that are not very close to the
cutoffs.  Without $S$ the force would not be differentiable at the cutoff
distances and hence a noticeable violation of energy conservation would
result in microcanonical Molecular Dynamics (MD) runs \cite{r38,r39}.

\begin{figure}
\includegraphics[clip,width=12.5cm]{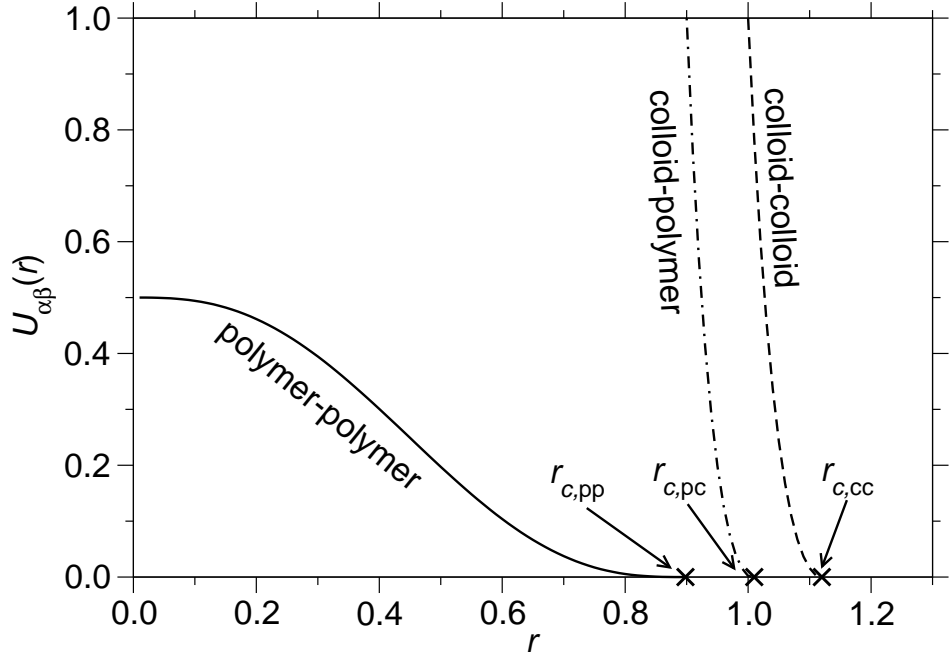}
\caption{Polymer-polymer, colloid-polymer, and colloid-colloid
interaction plotted as a function of distance. Note that
energies $U_{\alpha\beta}(r)$ are measured in units of
$\epsilon_{\cc\cc}=\epsilon_{\cc\pp}=\kb T=1$.
\label{fig:1}}
\end{figure}

For the soft polymer-polymer potential the following somewhat arbitrary
but convenient choices are made:
\begin{equation}\label{eq:4}
U_{\pp\pp}(r)=  8\epsilon_{\pp\pp} \biggl [ 1
-10\left(\frac{r}{r_{c,{\pp\pp}}}\right)^3
+15\left(\frac{r}{r_{c,{\pp\pp}}}\right)^4
-6\left(\frac{r}{r_{c,{\pp\pp}}}\right)^5  \biggr]\,,
\end{equation}
where $r_{c,\pp\pp}=2^{1/6}\sigma_{\pp\pp}(=0.8r_{c,{\cc\cc}})$ and
\begin{align}
&\text{(i)} & \epsilon_{\pp\pp}&=0 &&\text{(soft AO model)} \label{eq:5}\\
\text{or}& \nonumber \\
&\text{(ii)}& \epsilon_{\pp\pp}&=0.0625&\;&\text{(interacting polymers).} 
\label{eq:6}
\end{align}
Note that  the expansion in the square bracket of Eq.~\eqref{eq:4}
is essentially a polynomial fit to a cosine function, which is
shifted by unity and the angle of which varies from $0$ to $\pi$
when $r$ increases from zero to $r_{c,\pp\pp}$.  However, while
the cosine function is smoothly differentiable only once at $r=0$
and $r=r_{c,{\pp\pp}}$, Eq.~\eqref{eq:4} is twofold differentiable.
Of course, $U_{\pp\pp}(r>r_{c,{\pp\pp}})=0$.  Note that the choice
\eqref{eq:5} differs from the original AO model only by replacing the
original hard core interactions $U_{\cc\cc}(r)$ and $U_{\cc\pp}(r)$ by
smooth interactions, Eqs.~\eqref{eq:ljpotential}-\eqref{eq:ljsigma},
while polymers are still strictly non-interacting.  For the choice
\eqref{eq:6}, which is the choice used for the MD work, the energy
varies from $U_{\pp\pp}(r=0)=1/2\,\kb T$ to zero, Fig.~\ref{fig:1}.
With these choices of potentials the application of MD is straightforward
and efficient, but also the application of grand-canonical Monte Carlo
methods is still well feasible.

However, as in our earlier study of static and dynamic critical phenomena
of a symmetrical binary Lennard-Jones mixture \cite{r42,r43,r44,r45} it is
advantageous to apply Monte Carlo methods (in the fully grand-canonical
$(N\mu_{\cc}\mu_{\pp}T)$-ensemble, with $\mu_{\cc}$, $\mu_{\pp}$ being
the chemical potentials of colloids and polymers, respectively, in the
present case) to determine the static phase diagram of the colloid-polymer
mixture.  In particular, determining the critical densities of colloids
$\rho_{\cc}^\crit$ and polymers $\rho_\pp^\crit$ (where densities are
defined in terms of the particle numbers of colloids $N_\cc$ and polymers
$N_\pp$ in the standard way $\rho_\cc=N_\cc/V$, $\rho_\pp=N_\pp/V$, $V$
being the volume of the simulation box) is a nontrivial matter.  In the
context of MD simulation, starting systems at states which fall within the
two-phase coexistence region and beginning with an initially homogeneous
distribution of both types of particles lead to a phase separation into
a ``liquid-like'' phase (with densities $\rho_\cc^l$, $\rho_\pp^l$)
and a ``vapor-like'' phase (with densities $\rho_\cc^v$, $\rho_\pp^v$).
Of course, a priori all the values of densities along the coexistence
curve in the $(\rho_\cc,\rho_\pp)$-plane are unknown and simulating the
dynamics of spinodal decomposition is a complicated and notoriously slow
process \cite{r46,r47}.  Moreover, when approaching the critical point
(from the one-phase region or along the coexistence curve) simulations
in the canonical ensemble suffer severely from critical slowing down
\cite{r48} as discussed in \cite{r44,r45}.

\begin{figure}
\includegraphics[clip,width=12.5cm]{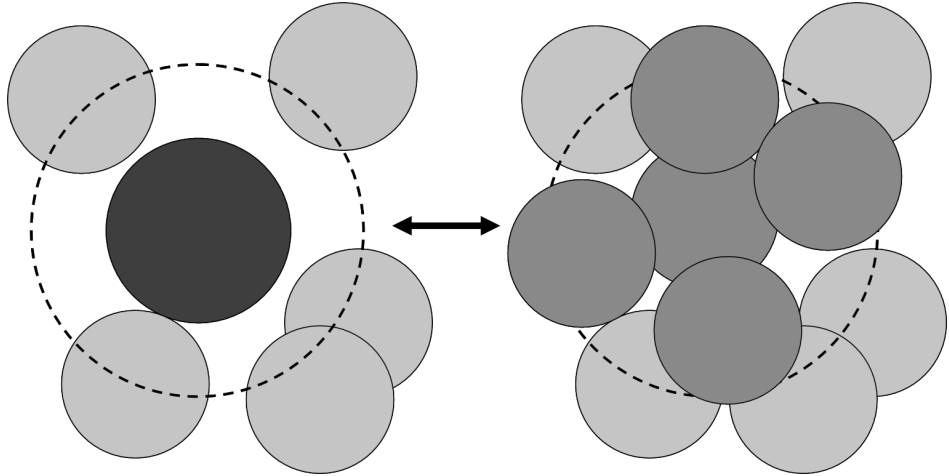}
\caption{Schematic illustration of the cluster move applied for the
grand-canonical Monte Carlo simulation. One colloidal particle (black
sphere) is replaced by several overlapping polymer particles (dark grey)
with randomly chosen center of mass positions inside the depletion zone
of the colloid (indicated by the dashed circle). Other polymers in the
environment are shown as light grey spheres. The double arrow indicates
that the inverse move is implemented as well.
\label{fig:2}}
\end{figure}

Thus, it is very desirable to study the phase behavior by Monte Carlo
simulations in the grand-canonical ensemble, which turned out to be very
useful for both the symmetrical binary Lennard-Jones mixture \cite{r44}
and the standard AO model \cite{r19a,r19b}.  %Note that when using
a number of well-equilibrated system configurations as initial states
for strictly microcanonical MD runs \cite{r42,r43,r44,r45}, one realizes
averages corresponding to a well-defined temperature $T$ without the need
to augment the MD code by a thermostat \cite{r38,r39}.  However, already
for the standard AO model straightforward particle insertion Monte Carlo
moves, which are necessary to realize the grand-canonical ensemble, are
almost always rejected due to the large density of polymers $\rho_\pp$
in the system \cite{r19a,r19b}.  For the standard AO model Vink and
Horbach \cite{r19a,r19b} could cope with this difficulty by implementing
a cluster move.  A similar cluster move is used here (Fig.~\ref{fig:2}).
To maximize the efficiency of this algorithm, always all polymer particles
are removed in the depletion zone when a colloid is inserted.  The radius
of the depletion zone was $\sigma_{\cc\cc}+r_{c,\cc\pp}$.  At most $m=10$
polymers would be inserted or removed in one attempted cluster move.

When this algorithm is applied to soft potentials, slight modifications
of the implementation are required: Colloid deletion attempts must
always be rejected if any center of polymer particles is located in
the depletion zone.  Otherwise, colloid insertion and deletion moves
are no longer symmetric and detailed balance is violated.  Note that
this problem does not occur in the original AO model because polymer
particles can never ``overlap'' with colloid particles.  Note that the
algorithm is still ergodic, because ``overlaps'' with the colloidal
particles can be obtained by removing adjacent colloids and filling the
void with polymer particles.  Nevertheless it is still recommendable to
mix cluster moves with local moves.

We wish to compare our results with the original AO model (with hard
core interactions), where it is standard practice to use the packing
fractions $\eta_\cc, \eta_\pp$ as variables,
\begin{equation}
\eta_\cc=\rho_\cc V_\cc\,,\quad\eta_\pp=\rho_\pp V_\pp\,,
\end{equation}
where $V_\cc$ and $V_\pp$ are the volumes occupied by a colloid and
polymer, respectively, with $V_\cc=\pi d^3_{\cc\cc}/6$ and $V_\pp=\pi
d^3_{\pp\pp}/6$, $d_{\cc\cc}$ and $d_{\pp\pp}$ being the diameters of
colloids and polymers.  For this comparison it is hence useful to define
an effective diameter of colloids and polymers of our model using the
approach of Barker and Henderson \cite{r49}
\begin{equation}\label{eq:barkerhenderson}
d_{\alpha\beta}=\int_0^{\sigma_{\alpha\beta}} 
\left[1-e^{-\frac{U_{\alpha\beta}(r)}{\kb T}}\right]\text{d} r\,.
\end{equation}
Using Eqs.~\eqref{eq:ljpotential}, \eqref{eq:smoothing} in Eq.~\eqref{eq:barkerhenderson} yields
\begin{equation}
d_{\cc\cc}=1.01557\sigma_{\cc\cc}
\end{equation}
and $d_{\cc\pp}=0.9d_{\cc\cc}$.  We also use $d_{\pp\pp}=0.8d_{\cc\cc}$
and consequently derive the following formulas to convert our densities
into packing fractions
\begin{equation}
  \eta_\cc=0.54844\sigma^3_{\cc\cc}\rho_\cc   \,,\quad   \eta_\pp=0.28080\sigma^3_{\cc\cc}\rho_\pp\,.
\end{equation}
Hence, the polymer reservoir packing fraction \cite{r10,r19a,r19b} is
given by $\eta^r_\pp=V_\pp \exp(\mu_\pp/\kb T)=0.28080\, \sigma^3_{\cc\cc}
\exp(\mu_\pp/\kb T)$ in terms of the chemical potential $\mu_\pp$ of the
polymers.  Of course, for interacting polymers the notion of $\eta^r_\pp$
loses its original meaning, but we continue to use $\eta^r_\pp$ as
defined here for the sake of comparability with the standard AO model.

\begin{figure}
\includegraphics[clip,width=12.5cm]{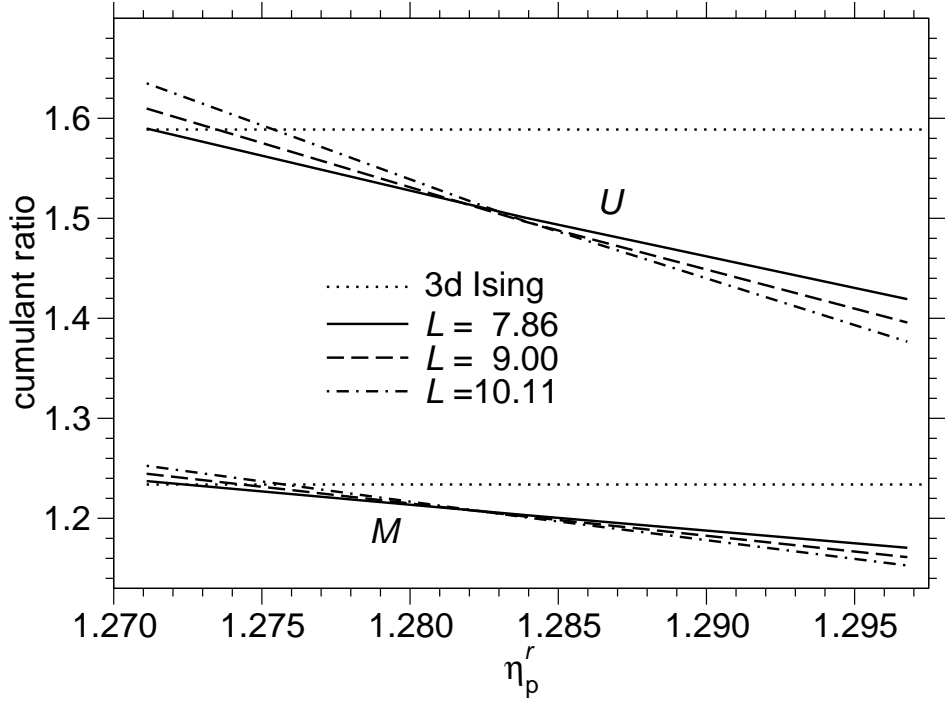}
\caption{Cumulant ratios $M$ and $U$ as a function of polymer reservoir
packing fraction $\eta_\pp^r$ for the model with interacting polymers
(with $8\epsilon_{\pp\pp}=0.5$). Three linear dimensions $L$ are included
($L$ is measured in units of $\sigma_{\cc\cc}$). The horizontal lines
indicate the universal values \cite{r56} $M\approx 1.239$ and $U\approx
1.589$ which $M$ and $U$ should acquire at criticality for every system
in the universality class of the three-dimensional Ising model.
\label{fig:3}}
\end{figure}

The technical aspects of grand-canonical Monte-Carlo simulations of
phase equilibria and critical phenomena in colloid-polymer mixtures have
been described in detail in the literature \cite{r19a,r19b,r20,r23}.
Therefore, we recall only very briefly the most salient features.
Choosing $\mu_\pp$ and hence $\eta^r_\pp$ as a parameter, the chemical
potential $\mu_\cc$ of the colloids is varied and the distribution
$P(\eta_\cc)$ of the colloid volume fraction is sampled, applying the
cluster algorithm mentioned above.  For $\eta^r_\pp$ sufficiently less
than the critical volume $\eta^r_{\pp,\crit}$ (note that $\eta^r_\pp$
plays the role of inverse temperature, when the phase diagram of
the colloid-polymer mixture is compared to the vapor-liquid phase
separation of a molecular system) $P(\eta_\cc)$ has a single peak
at $\langle\eta_\cc\rangle$ and the task is straightforward.  For
$\eta_\pp^r>\eta^r_{\pp,\crit}$ and $\mu$ near $\mu_\text{coex}$, however,
$P(\eta_\cc)$ is a doubly-peaked function where (apart from finite
size effects \cite{r19a,r19b,r20,r50,r51,r52}) the positions of the two
peaks correspond to the volume fractions of the vapor-like phases at the
coexistence curve, $\eta_\cc^v(\eta_\pp^r)$ and $\eta_\cc^l(\eta_\pp^r)$.
Since the distribution $P(\eta_\cc)$ develops a very deep minimum in
between these peaks \cite{r52,r53}, sampling is, however, not completely
straightforward.  An efficient way to overcome this difficulty is provided
by \emph{successive umbrella sampling} \cite{r54}, which was applied in
this work.  The chemical potential $\mu_\text{coex}$ at vapor-liquid
coexistence is then given by the \emph{equal weight rule} \cite{r55},
i.e.~the areas underneath the peaks corresponding to the vapor-like phase
and the liquid-like phase have to be equal.  The order parameter $m$
of the phase transition can then be identified as
\begin{equation}
m=\eta_\cc-\langle\eta_\cc\rangle\,, \label{eq:11}
\end{equation}
with $\langle\eta_\cc\rangle$ being the average of $P(\eta_\cc)$ including
both peaks at $\mu_\text{coex}$.  A convenient tool to find the critical
value $\eta^r_{\pp,\crit}$ is based on the analysis of moment ratios $M$,
$U$ defined as \cite{r50,r52}
\begin{equation}
M=\frac{\langle m^2\rangle}{\langle |m|\rangle ^2}\,,\quad 
U=\frac{\langle m^4\rangle}{\langle m^2\rangle ^2}\,,
\end{equation}
while following a path along $\mu_\text{coex}(\eta^r_\pp)$ for different
linear dimensions $L$ of the cubic simulation box.  The critical value
$\eta^r_{\pp,\crit}$ is determined found from the intersection of these
curves.  Figure \ref{fig:3} gives an example for the present model.
Note that the procedure described above is also operationally well
defined for a range of values $\eta_\pp^r<\eta^r_{\pp,\crit}$, since due
to finite size effects, the distribution $P(\eta_\cc)$ is double-peaked
over some range in the one-phase region as well \cite{r50,r52}.  It can be
recognized from Fig.~\ref{fig:3} that a rather well-defined intersection
point occurs for $\eta^r_{\pp,\crit}=1.282\pm0.002$.  However,
this intersection does not occur at the theoretical value \cite{r56}
$M\approx 1.239$ but at a somewhat lower value $M_\text{eff}\approx 1.21$.
This discrepancy is due to various corrections to finite size scaling,
in particular the so-called \emph{field mixing effects} \cite{r57a,r57b}.
In the case of the standard AO model, a very similar discrepancy occurs
as well \cite{r19a,r19b}.  Since in the latter model no potential energy
is present, the field mixing does not involve a coupling between energy
density and density as for ordinary fluids \cite{r57a,r57b} but rather
a coupling between colloid density and polymer density.  So the order
parameter (in the sense of a \emph{scaling field} \cite{r57a,r57b})
is in a strict sense not given by $\eta_\cc$ alone (as assumed in
Eq.~\eqref{eq:11}).  Instead, a suitable linear combination of $\eta_\cc$
and $\eta_\pp$ needs to be constructed.  However, we have not done
this in the context of finding the critical point since there is ample
evidence in various systems \cite{r58,r59,r60} that the simple cumulant
intersection method as illustrated in Fig.~\ref{fig:3} does yield the
critical point with a relative accuracy of a few parts in a thousand,
which suffices for the present purposes.

\section{Static properties of the soft version of the AO model}\label{sec:statics}
\begin{figure}
\includegraphics[clip,width=12.5cm]{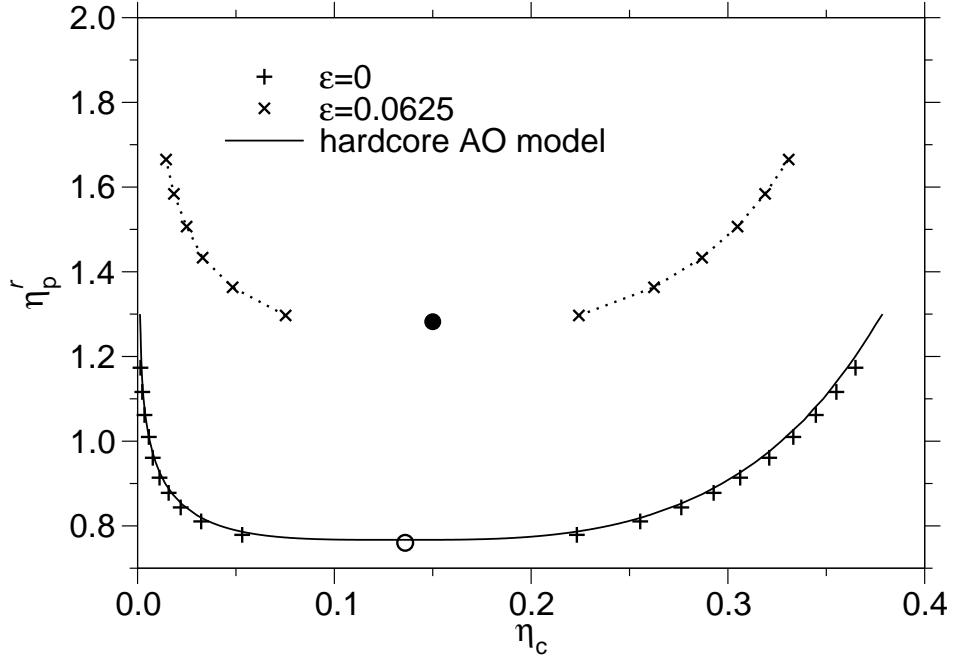}
\caption{Coexistence curves of the hard core AO model (from
\cite{r19a,r19b}, full curve), the soft AO model \eqref{eq:5}, and the
model with interacting polymers \eqref{eq:6} in the plane of variable
$\eta_\cc$, $\eta_\pp^r$ (reservoir representation). The open circle
marks the locus of the critical point for the hard core and the soft AO
model with $\epsilon=0$. The full dot shows the critical point for the
model with interacting polymers ($\epsilon=0.0625$).
\label{fig:4}}
\end{figure}

As discussed in Sec.~\ref{sec:model}, the first step of the Monte Carlo
study consists of the estimation of the coexistence curve and the critical
point.  For the two models defined in Eqs.~\eqref{eq:5},\eqref{eq:6}
we found for $\epsilon_{\pp\pp}=0$
\begin{equation}
\eta^r_{\pp,\crit}=0.760 \,,\quad \eta_{\cc,\crit}=0.136 \,,\quad  
\eta_{\pp,\crit}=0.354 
\end{equation}
and for $\epsilon_{\pp\pp}=0.0625$
\begin{equation} \label{eq:14}
\eta^r_{\pp,\crit}=1.282  \,,\quad  
\eta_{\cc,\crit}=0.150  \,,\quad  
\eta_{\pp,\crit}=0.328 \,.   
\end{equation}
Since the accuracy of these numbers is about $\pm 0.002$, we conclude that
model (i), the ``soft AO model'', is within our errors not distinguishable
from the original AO model with hard core interactions for which the
analogous results are \cite{r19a,r19b}
\begin{equation}
\eta^r_{\pp,\crit}=0.766  \,,\quad  
\eta_{\cc,\crit}=0.134  \,,\quad  
\eta_{\pp,\crit}=0.356  \,.
\end{equation}
This coincidence between the soft AO model and its hard core version
is also seen in the coexistence curve, which is compared in reservoir
representation in Fig.~\ref{fig:4}.  The coexistence curve of the model
with interacting polymers is substantially different, of course, as
expected from Eq.~\eqref{eq:14}.

\begin{figure}
\includegraphics[clip,width=12.5cm]{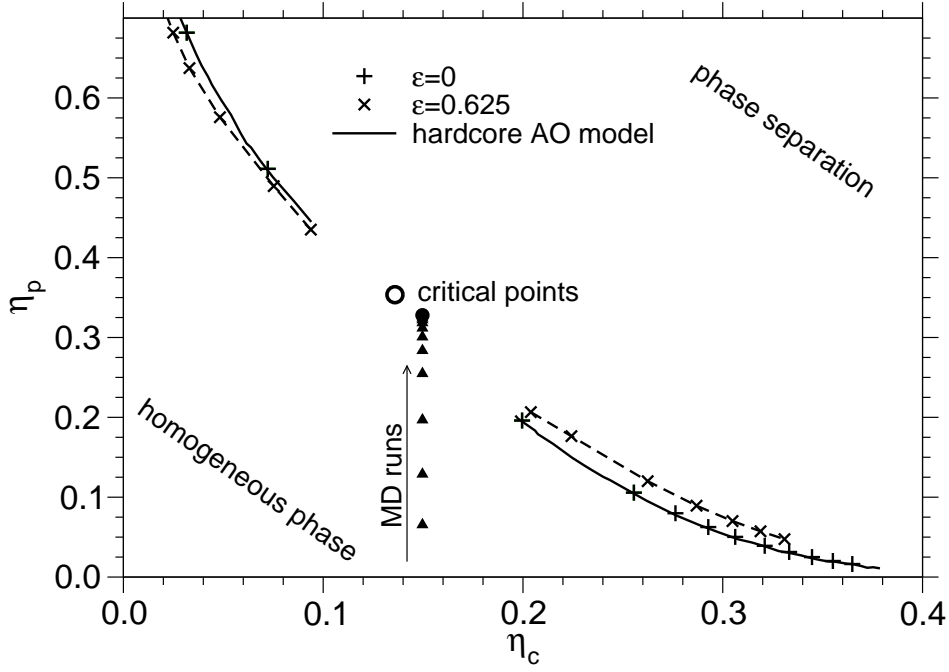}
\caption{Phase diagrams of colloid-polymer mixture models in the plane of
variables $\eta_\pp$ (polymer packing fraction) and $\eta_\cc$ (colloid
packing fraction). The original AO model (full curve), the soft AO model
(standing crosses), and the model with interacting polymers (dashed curve)
are compared. The triangles indicate state points at which MD runs were
performed. Each coexistence simulation took 24 hours on a 32-core Power4
cluster (1.7 GHz).
\label{fig:5}}
\end{figure}

However, when we study phase coexistence as a function of all
experimentally accessible variables $\eta_\pp, \eta_\cc$, differences
between the three models are rather minor (Fig.~\ref{fig:5}).
It appears that near criticality the main effect of ``switching
on'' the polymer-polymer interaction is to shift the critical point
along the coexistence curve of the AO model to the higher value of
$\eta_{\cc,\crit}$ (and correspondingly lower value of $\eta_{\pp,\crit}$
mentioned in Eq.~\eqref{eq:14}.  Further away from the critical point
the coexistence curve of the interacting polymer model predicts somewhat
lower polymer packing fractions along the ``vapor branch'' and somewhat
higher polymer packing fractions along the ``liquid branch''.  The result
that the critical packing fraction of polymers is about twice that of
the colloids is similar to what was observed in a recent experiment
\cite{r15}.  Note however, that in these experiments a size ratio of
polymers to colloids of $q=1.04$ (rather than $q=0.8$) was employed
which affects $\eta_{\cc,\crit}$ ($\eta_{\cc,\crit}\approx 0.10$ was
found in \cite{r15}).

In Fig.~\ref{fig:5} we have also indicated the state points at
which micro-canonical MD runs took place.  We have fixed the
number of colloids $N_\cc$ in a volume (of size $27^3$) such that
$\eta_\cc=\eta_{\cc,\crit}\approx 0.15$ (which corresponds to
$N_\cc=5373$).  Then the number of polymers was varied from zero up
to $N_\pp=22734$.  The MD runs were carried out with the Velocity
Verlet algorithm \cite{r38,r39} and a time step $\delta t=0.0005
(\sigma_{\cc\cc}^2 m_\cc / \epsilon_{\cc\cc})^{1/2}$.  The masses
of colloids and polymers are equal and units of time are chosen such
that $m_\cc=m_\pp=1$.  In the production runs used for the computation
of time-displaced correlation functions no thermostat was applied, so
a microcanonical ensemble respecting all conservation laws applies.
Starting configurations were generated as follows: First, a random
configuration was generated in a box of linear dimension $L=9$
and periodic boundary conditions.  The system was equilibrated at
$T=1$ for $20$ million time steps with a simple velocity rescaling
according to the Maxwell-Boltzmann distribution.  Then, the system
is enlarged from $L=9$ to $L=27$ by replicating it three times in
all spatial directions.  Now periodic boundary condition for $L=27$
only are applied.  Equilibration is continued for 2 million time steps,
again with a Maxwell-Boltzmann thermostat.  During this equilibration,
the original periodicity with $L=9$ is quickly lost.  The production
runs for static averages are done without applying any thermostat.
First, 5 million time steps are performed during which (at eight
different times) statistically independent configurations are stored.
These serve as starting configurations for eight independent simulation
runs, each with 5 million steps, for the computation of static averages.
During each run, 500 configurations are analyzed in regular intervals.
Thus, 4000 statistically independent configurations are averaged over
for the computation of the structure factor.

\begin{figure}
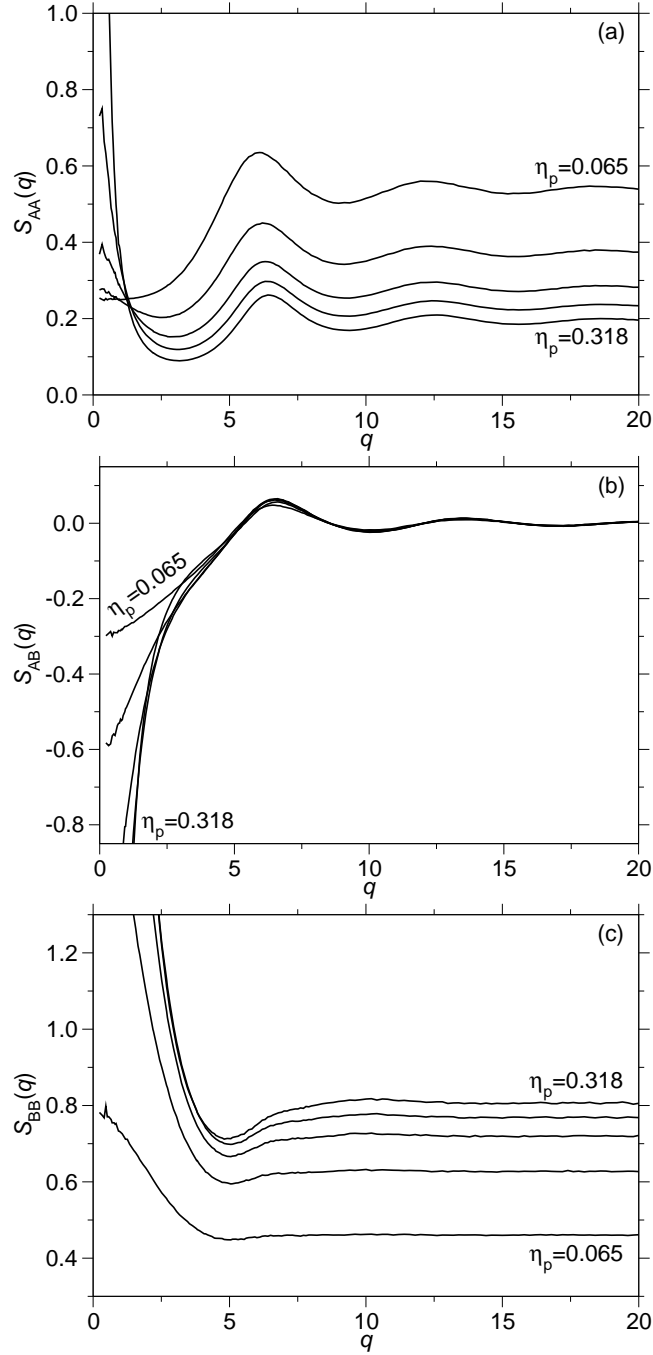

\includegraphics[clip,width=8.5cm]{fig6a}
\includegraphics[clip,width=8.5cm]{fig6b}
\includegraphics[clip,width=8.5cm]{fig6c}
\caption{(a) Partial structure factor $S_{\rma\rma}(q)$ describing the
scattering from colloids only, choosing $\eta_\cc=\eta_{\cc,\crit}=0.150$
and various choices for $\eta_\pp$ as indicated for a simulation box of
linear dimension $L=27$.  (b) Partial structure factor $S_{\rma\rmb}(q)$
describing the interference in the scattering from colloids and polymers.
(c) Partial structure factor $S_{\rmb\rmb}(q)$ describing the scattering
from polymers only.  The values for the polymer packing fractions
are $\eta_\pp=0.065, 0.129, 0.197, 0.255, 0.318$ and are also used in
Figs.~\ref{fig:7}-\ref{fig:8}.
\label{fig:6}}
\end{figure}

From now on, we denote colloids as A-particles and polymers as
B-particles.  For all simulated state points the partial structure
factors were computed,
\begin{equation}
S_{\alpha\beta}=\frac{1}{N}\left\langle\sum_{i=1}^{N_\alpha}
\sum_{j=1}^{N_\beta}
\exp(i\myvec{q}\cdot\myvec{r}_{ij})\right\rangle\,,\quad   
\alpha\in\rma, \rmb\,,
\end{equation}
with $N=N_\rma+N_\rmb$.  These results, presented in
Figs.~\ref{fig:6}(a)-(c), show that the partial structure factor for
colloids (Fig.~\ref{fig:6}(a)) displays an oscillatory structure with a
first peak near $q\approx 6.5$, which is typical for the packing of hard
particles in a moderately dense liquid.  The partial structure factor
due to polymers (Fig.~\ref{fig:6}(c)) exhibits much less structure in the
range of large $q$ as expected, since for the potential, Eq.~\eqref{eq:4},
the polymers can still overlap rather easily.  All these partial structure
factors show a strong enhancement at small $q$, reflecting the critical
scattering due to the unmixing tendency between colloids and polymers
when the critical point is approached.  Note that the partial structure
factor due to interference of the scattering from colloids and polymers
(Fig.~\ref{fig:6}(b)) also shows oscillations at large $q$, as does the
scattering from colloids alone (Fig.~\ref{fig:6}(a)).

\begin{figure}
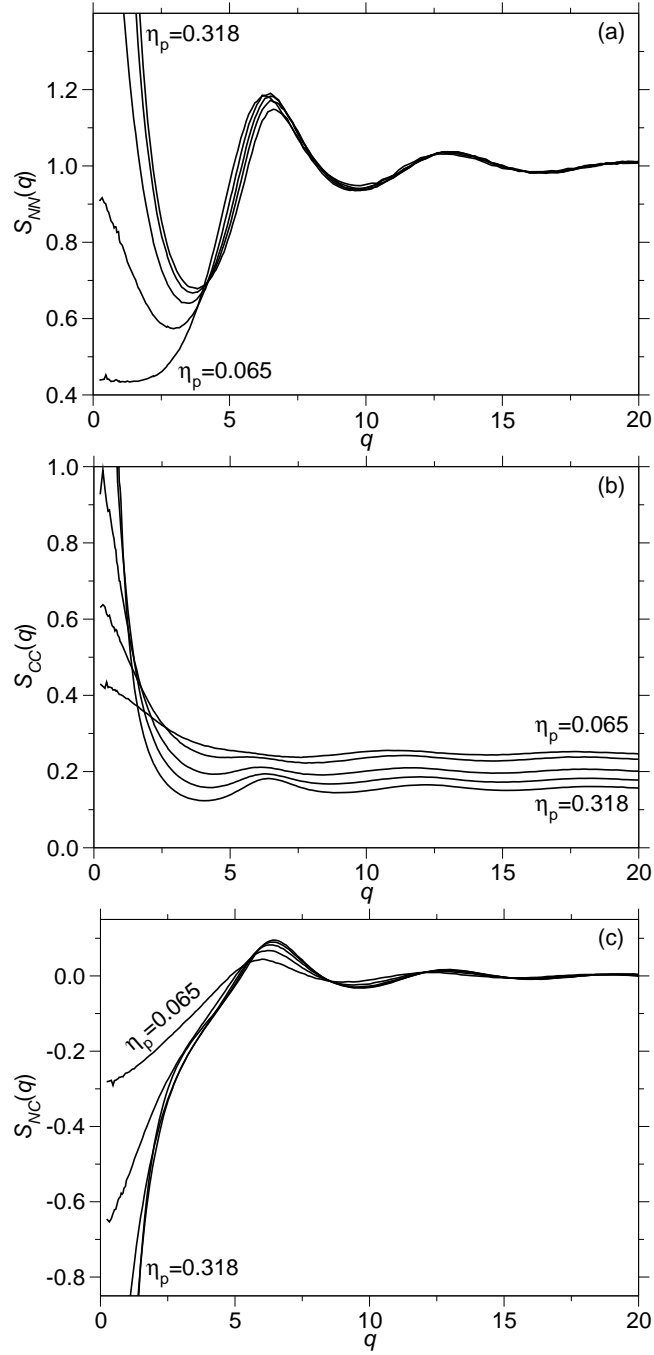

\includegraphics[clip,width=8.5cm]{fig7a}
\includegraphics[clip,width=8.5cm]{fig7b}
\includegraphics[clip,width=8.5cm]{fig7c}
\caption{(a) Number density structure factor $S_{NN}(q)$ calculated
from the data of of Fig.~\ref{fig:6}.  (b) Concentration structure
factor $S_{CC}(q)$.  (c) Density-concentration interference structure
factor $S_{NC}(q)$.
\label{fig:7}}
\end{figure}

From the partial structure factors it is useful to construct
combinations that single out number-density fluctuations $S_{NN}(q)$
and concentration fluctuations $S_{CC}(q)$, defined via \cite{r61}
($x_\rma=N_\rma/[N_\rma+N_\rmb]$, $x_\rmb=1-x_\rma$)
\begin{align}
S_{NN}(q)  & = S_{\rma\rma}(q)+2S_{\rma\rmb}(q)+S_{\rmb\rmb}(q)\,,\\
S_{CC}(q)  & = x_\rmb^2 S_{\rma\rma}(q)+x_\rma^2S_{\rmb\rmb}(q)
-2x_\rma x_\rmb S_{\rma\rmb}(q)\,.
\end{align}
In addition, it is of interest to consider a structure factor relating to
the coherent interference of number density and concentration fluctuations
\cite{r61},
\begin{equation}
S_{NC}(q) = x_\rmb S_{\rma\rma}(q)-x_\rma S_{\rmb\rmb}(q)
+(x_\rmb-x_\rma)S_{\rma\rmb}(q)\,.
\end{equation}
Figure \ref{fig:7} shows that all three structure factors show a
strong increase at small $q$, reflecting the critical scattering as
the critical point is approached.  Additionally, at large $q$ they
display oscillations.  The behavior seen in Fig.~\ref{fig:7} differs
very much from the behavior found for the unmixing of the symmetric
binary Lennard-Jones mixture \cite{r42,r43,r44,r45}.  In the latter case
$S_{NN}(q)$ was not sensitive to the critical fluctuations at all, which
showed up in $S_{CC}(q)$ only.  Likewise, $S_{CC}(q)$ was insensitive to
the way how the particles are ``packed'' in the liquid, i.e.~there was
no structure at large $q$.  In addition, almost no interference between
the scattering from concentration and density fluctuations could be seen.
Hence, $S_{NC}(q)$ was very small, while in the present model $S_{CC}(q)$
and $S_{NC}(q)$ are of the same order of magnitude.  These observations
clearly show that neither the total density in the system, nor the
relative concentration of one species is a ``good'' order parameter
of the phase separation that occurs.  (Likewise, Fig.~\ref{fig:6}
shows that neither the colloid density alone nor the polymer density
alone are ``good'' order parameters since both densities reflect the
critical scaling in a similar way.)  Of course, from the phase diagram
(Fig.~\ref{fig:5}) such a problem is expected since the shape of the
coexistence curve shows that the order parameter is a nontrivial linear
combination of both particle numbers $N_\pp, N_\cc$.

\begin{figure}
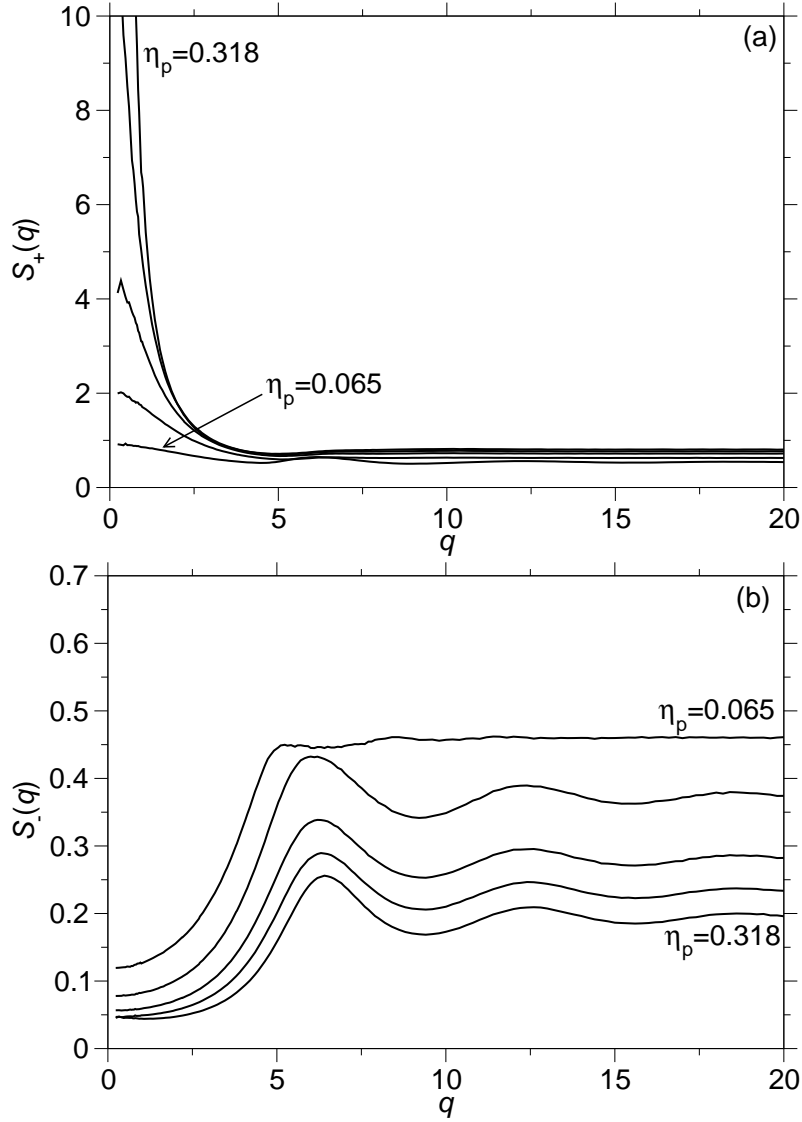

\includegraphics[clip,width=10.5cm]{fig8a}
\includegraphics[clip,width=10.5cm]{fig8b}
\caption{Structure factors $S_+(q)$, (a), and $S_-(q)$, (b), plotted
versus $q$ for several values of $\eta_\pp$ as indicated.  $S_+(q)$
and $S_-(q)$ are the two eigenvalues of the matrix $\mymatrix{S}(q)$,
cf.~Eqs.~\eqref{eq:n20},\eqref{eq:n21}.
\label{fig:n8}}
\end{figure}

\begin{figure}
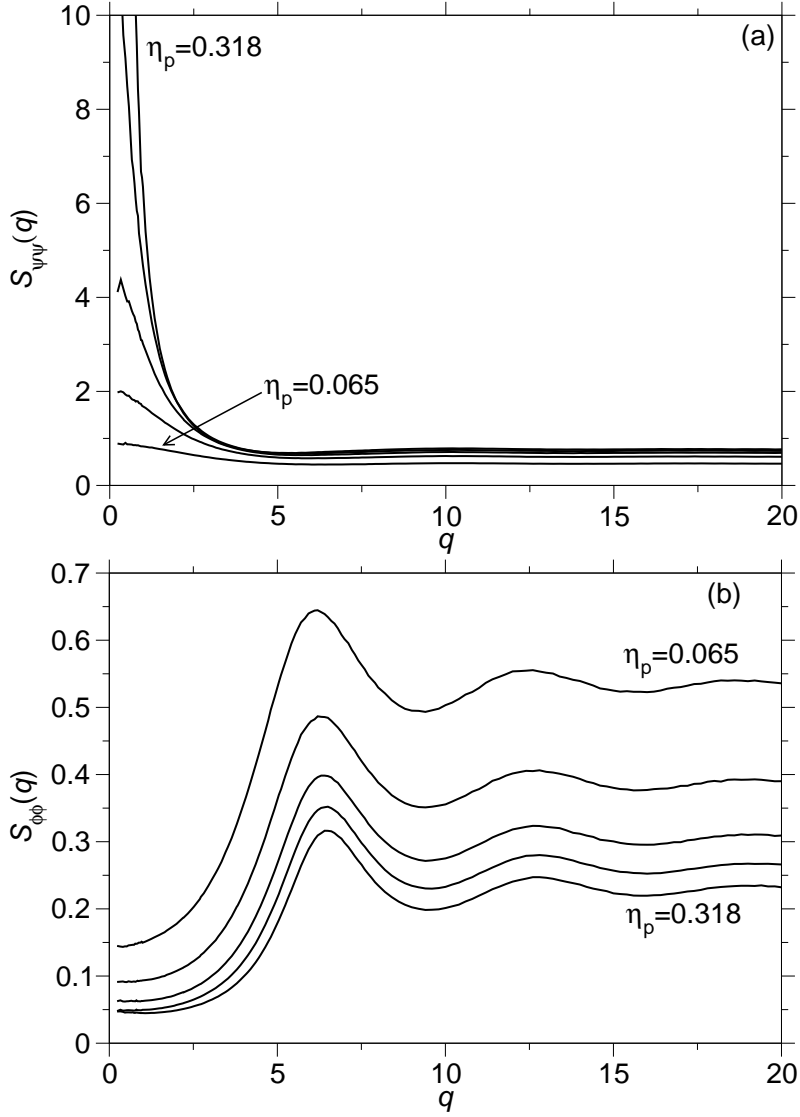

\includegraphics[clip,width=10.5cm]{fig9a}
\includegraphics[clip,width=10.5cm]{fig9b}
\caption{Structure factors $S_{\psi\psi}(q)$, (a), and $S_{\phi\phi}(q)$,
(b), plotted versus $q$ for the same values of $\eta_\pp$ as in
Fig.~\ref{fig:n8}.  Coefficients are $a=-0.24, b=0.97, a'=-0.97,
b'=-0.24$.
\label{fig:n9}}
\end{figure}

In order to deal with this problem, we introduce a symmetrical matrix
formed from the structure factors $S_{\rma\rma}(q)$, $S_{\rma\rmb}(q)$
and $S_{\rmb\rmb}(q)$
\begin{equation} \label{eq:n20}
\mymatrix{S}(q)=\begin{pmatrix}
S_{\rma\rma}(q)   &  S_{\rma\rmb}(q) \\
S_{\rma\rmb}(q)   &  S_{\rmb\rmb}(q)
\end{pmatrix}
\end{equation}
and diagonalize this matrix to obtain its diagonal form
\begin{equation} \label{eq:n21}
\mymatrix{S}^{(d)}(q)=\begin{pmatrix}
S_+(q)   &  0 \\
0        &  S_-(q)
\end{pmatrix}   \,,
\end{equation}
with
\begin{equation}
S_{\pm}(q)=\frac{1}{2}[S_{\rma\rma}(q)+S_{\rmb\rmb}(q)]   \pm
\sqrt{\frac{1}{4}[S_{\rma\rma}(q)-S_{\rmb\rmb}(q)]^2+
S_{\rma\rmb}^2(q)}   \,.
\end{equation}
Figure \ref{fig:n8} shows a plot of $S_+(q)$ and $S_-(q)$ versus $q$.
This plot shows that this procedure indeed resulted in a decoupling of
the order parameter fluctuations (which show a critical enhancement
as $q\to 0$), as being measured by $S_+(q)$, and the noncritical
``particle packing'' fluctuations, measured by $S_-(q)$, which show
the characteristic oscillatory structure of a noncritical fluid.
In the case of the symmetrical LJ mixture the transformation
from the number density fluctuations of A and B particles to the
structure factors measuring the fluctuations of the total density of
particles and of their relative concentrations is unambiguous.  In the
case of the colloid-polymer mixture it is none of these variables
which plays the role of an order parameter, but a different linear
combination of both local densities of A and B particles, related to
the eigenvector corresponding to $S_+(q)$.  We can give this fact a
plausible interpretation by constructing two linear combinations of
the operators $\rho_\rma(\myvec{q})$, $\rho_\rmb(\myvec{q})$, defined via
$\rho_\alpha(\myvec{q})=\sum_{i=1}^{N_\alpha}\exp(i\myvec{q}\cdot\myvec{r}_\alpha)$,
as follows
\begin{align}
\psi(\myvec{q})&=a  \rho_\rma(\myvec{q}) + 
b \rho_\rmb(\myvec{q})\,,\label{eq:n22}\\
\phi(\myvec{q})&=a' \rho_\rma(\myvec{q}) + 
b' \rho_\rmb(\myvec{q})\,.\label{eq:n23}
\end{align}
The coefficients $a, b$ are defined such that at the critical point the
densities lie tangential to the coexistence curves.  Coefficients $a', b'$
are chosen such that the densities vary in a perpendicular direction to
this slope.  When we construct the structure factors (Fig.~\ref{fig:n9})
\begin{equation}\label{eq:n24}
S_{\psi\psi}(q)=\frac{1}{N}\langle |\psi(q)|^2\rangle\,,\quad
S_{\phi\phi}(q)=\frac{1}{N}\langle |\phi(q)|^2\rangle\,,
\end{equation}
one recognizes that $S_{\psi\psi}(q)$ is very similar to $S_+(q)$
and $S_{\phi\phi}(q)$ very similar to $S_-(q)$.  The structure factors
defined in this manner are not strictly identical to $S_+(q), S_-(q)$.
With increasing distance from criticality the relative weights $b/a$,
$b'/a'$ of the components of the ``order parameter components'' $\psi(q),
\phi(q)$ change.

\begin{figure}
\includegraphics[clip,width=12.5cm]{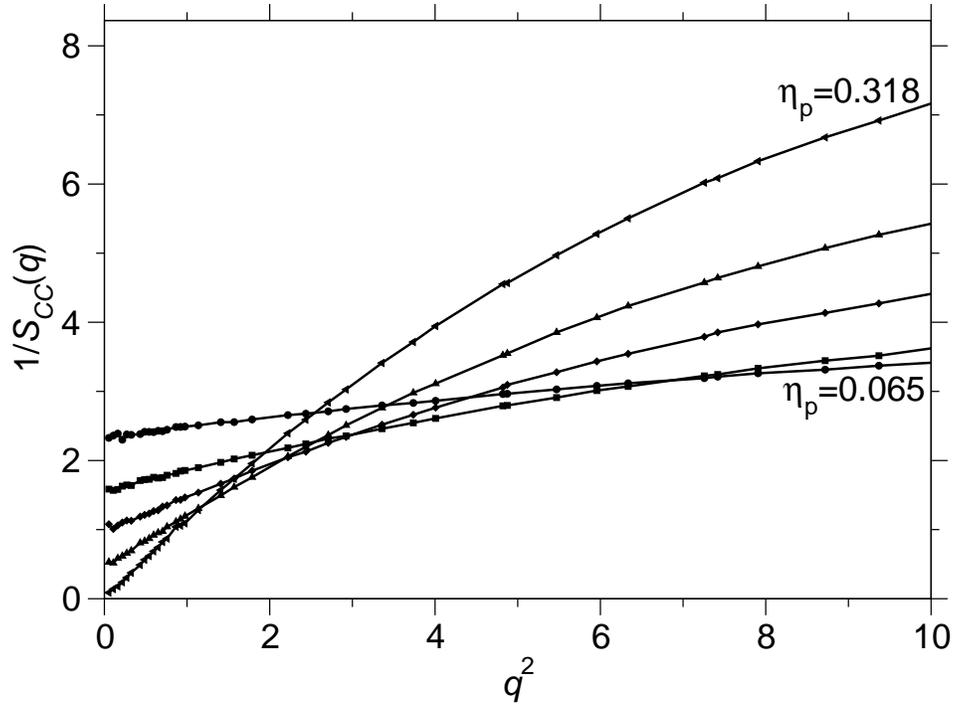}
\caption{Plot of the inverse of the concentration fluctuation structure
factor $1/S_{CC}(q)$ versus $q^2$. In the range $0<q^2<2$ data points
are fitted to the Ornstein-Zernike relation \eqref{eq:20}.
\label{fig:8}}
\end{figure}

\begin{figure}
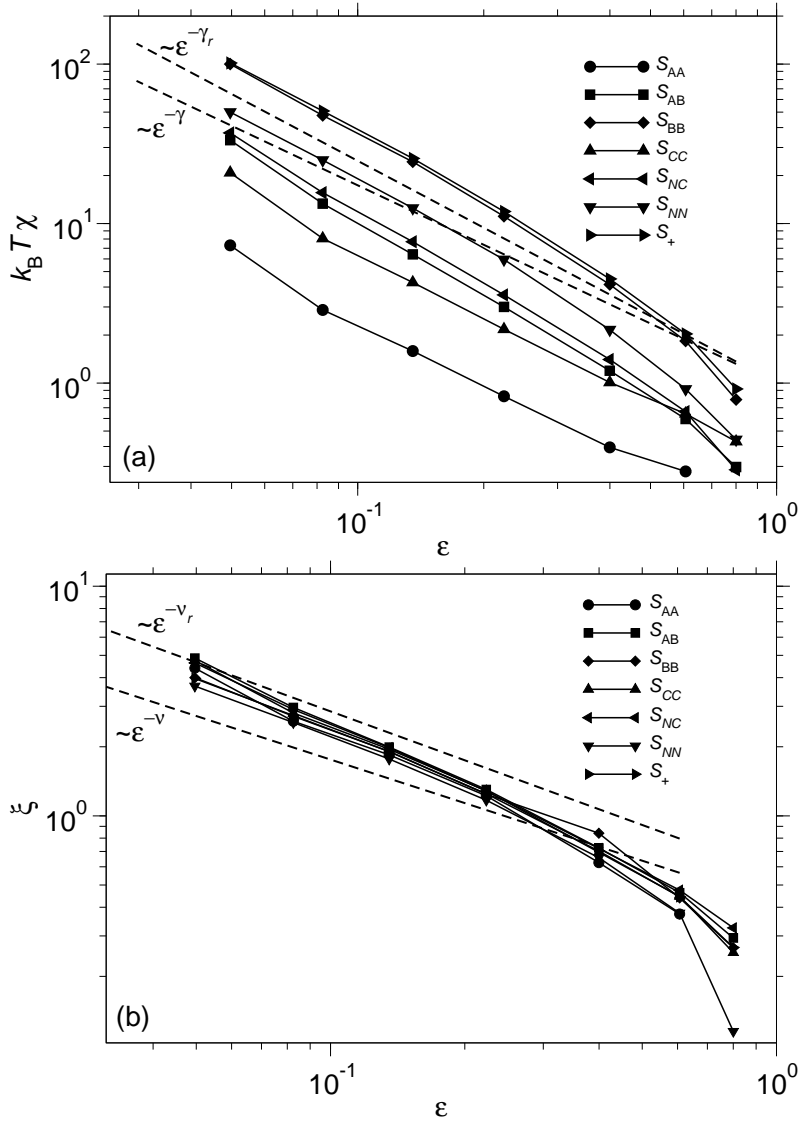

\includegraphics[clip,width=10.5cm]{fig11a}
\includegraphics[clip,width=10.5cm]{fig11b}
\caption{Log-log plots of (a) $\kb T \chi$ and (b) $\xi$ versus
$\epsilon=1-\eta_\pp/\eta_{\pp,\crit}$ from MD simulations.  Dashed and
dashed-dotted lines indicate power law fits with (a) the exponents $\kb
T\chi \propto \epsilon^{-\gamma}$ and $\epsilon^{-\gamma_r}$ and (b)
$\xi\propto \epsilon^{-\nu}$ or $\epsilon^{-\nu_r}$, where $\gamma=1.24$
and $\nu=0.63$ are the standard Ising exponents \cite{r62a,r62b} while
$\gamma_r=\gamma/(1-\alpha)$ and $\nu_r=\nu/(1-\alpha)$ are the Fisher
renormalized exponents \cite{r63} where $\alpha\approx 0.11$ is the
critical exponent of the specific heat \cite{r63}.
\label{fig:9}}
\end{figure}

For $q\to 0$ all those structure factors that show a critical increase
can be described by the well-known Ornstein-Zernike behavior.  This is
illustrated in Fig.~\ref{fig:8}, as an example, for the concentration,
fitting $1/S_{CC}(q)$ versus $q^2$ at small enough $q$ ($q^2\ll 2$)
to the relation \cite{r41,r45}
\begin{equation}\label{eq:20}
S_{CC}^{-1}(q)=(\kb T\chi_{CC})^{-1}[1+q^2\xi_{CC}^2+\dots]\,,
\quad q\rightarrow 0\,.
\end{equation}
Here, $\chi_{CC}$ is the ``susceptibility'' describing the magnitude
of concentration fluctuations and $\xi_{CC}$ the correlation length.
The various susceptibilities relating to the various structure
factors defined above and the associated correlation ranges are
shown in Fig.~\ref{fig:9}.  It is gratifying to note that indeed the
``susceptibility'' related to $S_+(q)$ is the largest susceptibility
that can be found, while the estimates for the correlation lengths are
all equal (within statistical errors).  Due to the coupling between
variables, there is only a single correlation length in the problem.

In Fig.~\ref{fig:9} we have included two theoretical predictions in the
log-log plot for the critical exponents, one is a slope corresponding to
the standard Ising exponents (that are observed in the grand-canonical
ensemble, where only intensive thermodynamic variables are held constant).
The other slope shows the exponents if ``Fisher renormalization'' occurs.
To remind the reader of this phenomenon we note that the response function
$\chi\equiv (\partial N_\cc/\partial\mu)_{T,\mu_\pp}|_{\mu=\mu_\crit}/N$
that is observed via Monte Carlo from the fluctuation relation
\begin{equation}\label{eq:21}
\kb T \chi=N^{-1}(\langle N_\cc^2\rangle-\langle N_\cc\rangle^2)_{T,\mu_\pp}
\end{equation}
differs from $\chi_{CC}$ since fluctuations differ in different
ensembles of statistical mechanics.  While $\chi_{CC}$ was estimated from
Eq.~\eqref{eq:20} which refers to the ensemble where $N_\pp=\text{const}$,
Eq.~\eqref{eq:21} refers to the ensemble where $\mu_\pp=\text{const}$.
Since \cite{r63}
\begin{equation}\label{eq:22}
  N_\pp=N_{\pp,\crit}+a(\mu_\pp-\mu_{\pp,\crit})^{1-\alpha}+b(\mu_\pp-\mu_{\pp,\crit})+\dots\,,
\end{equation}
where $\alpha$ is the specific heat exponent and $a, b$ are constants.
Very close to the critical point we have a singular relation between
$\langle N_\pp\rangle-N_{\pp,\crit}$ and $\mu_\pp-\mu_{\pp,\crit}$, namely
\begin{equation}\label{eq:23}
\epsilon=\frac{N_\pp}{N_{\pp,\crit}}-1
\propto\left(\frac{\mu_\pp}{\mu_{\pp,\crit}}-1\right)^{1-\alpha}\,.
\end{equation}
Therefore the power laws in the grand-canonical ensemble \cite{r62a,r62b}
\begin{equation}
\chi\propto\left(\frac{\mu_\pp}{\mu_{\pp,\crit}}-1\right)^{-\gamma}\,,\quad
\xi\propto\left(\frac{\mu_\pp}{\mu_{\pp,\crit}}-1\right)^{-\nu}
\end{equation}
translate into power laws with ``Fisher renormalized'' \cite{r63}
exponents in the microcanonical ensembles where $N_\pp, N_\cc$ are
constant
\begin{equation}
\chi_{CC}\propto \epsilon^{-\gamma/(1-\alpha)}\,,\quad
\xi_{CC} \propto \epsilon^{-\nu   /(1-\alpha)}\,.
\end{equation}
However, since the regular third term on the right hand side of
Eq.~\eqref{eq:22} is comparable to the (singular) second term that
was only used in Eq.~\eqref{eq:23}, except if one works extremely
close to $\mu_{\pp,\crit}$, it is difficult to ascertain whether or
not the simulation data shows any signature of Fisher renormalization.
High precision simulations for very much larger systems would be required
to clearly resolve this issue.  This task, however, is not possible with
presently available computer resources.

It is also useful to recall that susceptibilities observed in the
grandcanonical ensemble differ from those extracted from structure factors
in the canonical ensemble.  To interpret this difference, we start from
the standard relation for the grandcanonical partition function
\begin{equation} \label{eq:n31}
Z_\text{gc} (\mu,\mu_\pp,V,T)=
\sum_{N_c=0}^\infty \exp\left(\frac{\mu N_\cc}{\kb T}\right)
\sum_{N_\pp=0}^\infty\exp\left(\frac{\mu_\pp N_\pp}{\kb T}\right) 
Z_\text{c}(N_\cc,N_\pp,V,T)
\end{equation}
from which one straightforwardly derives the following fluctuation
relations ($\rho_\cc=\langle N_\cc\rangle/V$):
\begin{align}
\kb T \frac{\partial \langle N_\cc\rangle}{\partial\mu}&=
\langle N_\cc^2\rangle -\langle N_\cc\rangle^2\,, & 
\frac{\kb T}{V} 
\frac{\partial \langle \rho_\cc\rangle}{\partial\mu}&=
\langle \rho_\cc^2\rangle -\langle \rho_\cc\rangle^2  \label{eq:n32}\\
\kb T \frac{\partial \langle N_\pp\rangle}{\partial\mu_\pp}&=
\langle N_\pp^2\rangle -\langle N_\pp\rangle^2\,, & 
\frac{\kb T}{V} \frac{\partial \langle \rho_\pp\rangle}{\partial\mu_\pp}&=
\langle \rho_\pp^2\rangle -\langle \rho_\pp\rangle^2  \label{eq:n33}
\end{align}
and
\begin{equation}\label{eq:n34}
\begin{split}
\kb T \frac{\partial \langle N_\cc\rangle}{\partial\mu_\pp}&=
\kb T \frac{\partial \langle N_\pp\rangle}{\partial\mu}=
\langle N_\cc N_\pp\rangle -\langle N_\cc\rangle\langle N_\pp\rangle\\
\text{or}\\
\frac{\kb T}{V} \frac{\partial \langle \rho_\cc\rangle}{\partial\mu_\pp}&=
\langle \rho_\cc \rho_\pp\rangle -\langle \rho_\cc\rangle\langle \rho_\pp\rangle
\end{split}
\end{equation}
It is this mixed susceptibility describing the correlations between the
fluctuations of colloid and polymer number which enters the difference
between the susceptibilities in the canonical and grandcanonical ensemble.
A simple calculation yields
\begin{align} \label{eq:n34_2}
\chi_{T,\rho_\pp}^\text{coll} & \equiv \frac{V}{N}
\left(\frac{\partial\langle\rho_\cc\rangle}{\partial\mu}\right)_{T,\rho_\pp} \nonumber\\
& =  \chi_{T,\mu_\pp}^\text{coll}-     \frac{V}{N}
\left[\frac{\partial\langle\rho_\cc\rangle}{\partial\mu_\pp}\right]^2
\left/  
\left(\frac{\partial\langle\rho_\pp\rangle}{\partial\mu_\pp}\right)_{\mu,T}
\right.\,.
\end{align}
Similarly,
\begin{align} \label{eq:n35}
\chi_{T,\rho_\cc}^\text{pol} & \equiv \frac{V}{N}
\left(\frac{\partial\langle\rho_\pp\rangle}{\partial\mu_\pp}\right)_{T,\rho_\cc} \nonumber\\
& =  \chi_{T,\mu}^\text{pol}-         \frac{V}{N}
\left[\frac{\partial\langle\rho_\cc\rangle}{\partial\mu_\pp}\right]^2
\left/  
\left(\frac{\partial\langle\rho_\cc\rangle}{\partial\mu}\right)_{\mu_\pp,T}
\right.\,.
\end{align}

\begin{figure}
\includegraphics[clip,width=12.5cm]{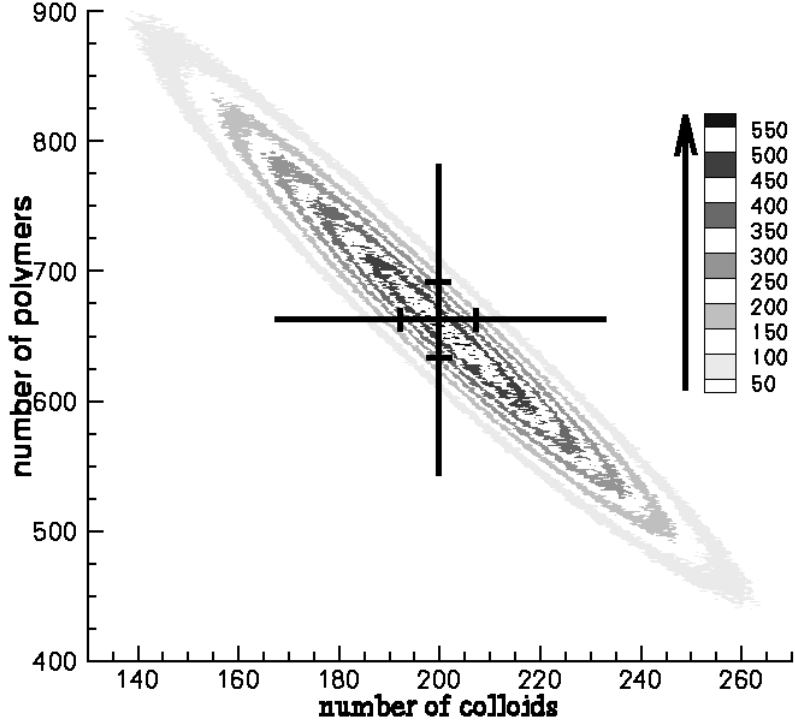}
\caption{Contour plot of a two dimensional probability distribution
$P(N_{\rm c}, N_{\rm p}$) in the one phase region ($\epsilon=0.0625$,
$\mu_{\rm c}=5.0148$, $\mu_{\rm p}=1.27973$, $L=9^{3}$.) The legend
describes the numbers of occurrence for each data point ($N_{\rm c}$,
$N_{\rm p}$). For better visibility data are grouped in bands. Note that
$x$- and $y$-axis have different scales.  $P(N_{\rm c}, N_{\rm p})$ can
be described as a bivariate Gaussian in $N_{\rm c}-\langle N_{\rm c}
\rangle$ and $N_{\rm p}-\langle N_{\rm p} \rangle$. Susceptibilities
$\chi_{T,\rho_{\rm p}}^{\rm coll}$ and $\chi_{T,\rho_{\rm c}}^{\rm
pol}$ can be extracted from the half-widths of these distributions
for $\rho_{\rm p}=\rangle \rho_{\rm p} \langle={\rm const}$ and
$\rho_{\rm c}=\langle \rho_{\rm c} \rangle={\rm const}$ (small black
bars), respectively. Similarly, $\chi_{T,\mu_{\rm p}}^{\rm coll}$
and $\chi_{T,\mu}^{\rm pol}$ can be obtained from the half-widths of
the projections to the $x$- and $y$-axis (large black bars). We can
also define an order parameter along the main axis of the ellipsoidal
contours which will maximize fluctuations and result in the order
parameter susceptibility $\chi_{+}$ as described in the text.
\label{fig:verteilung}}
\end{figure}

In fully grand-canonical simulations, as we have carried out in
the present work, it is possible to extract all susceptibilities of
interest from a study of the joint distribution function $P(N_\cc,
N_\pp)$.  In the one phase region and for large enough linear
dimensions $L$ this function is a bivariate Gaussian in the variables
$N_\cc - \langle N_\cc \rangle$, $ N_\pp - \langle N_\pp \rangle$,
see Fig.~\ref{fig:verteilung} for an explicit example.  Then the
fluctuations $\chi^\text{coll}_{T, \rho_\pp}$ and $\chi^\text{pol}_{T,
\rho_\cc}$ can be extracted from the half-widths of these distributions
along the abscissa direction ($\rho_\pp=N_\pp/V=\text{const}$) and
ordinate direction $(\rho_\cc=N_\cc/V=\text{const})$, respectively.
For the grand-canonical simulations described in Fig.12, for example,
we obtain $\chi_{T,\rho_{\rm p}=\langle \rho_{\rm p} \rangle}^{\rm
coll}=0.047$ and 0.05 for the second equation in (36) thus confirming
our calculations.  Figure \ref{fig:verteilung} illustrates again that
none of these susceptibilities should be regarded as the order parameter
susceptibility $\chi_+$: rather the latter is the half-width along the
main axis of the ellipsoidal contours $P(N_\cc, N_\pp)=\text{const}$
in Fig.~\ref{fig:verteilung}.

\section{Dynamics of colloid-polymer mixtures}\label{sec:dynamics}
\begin{figure}
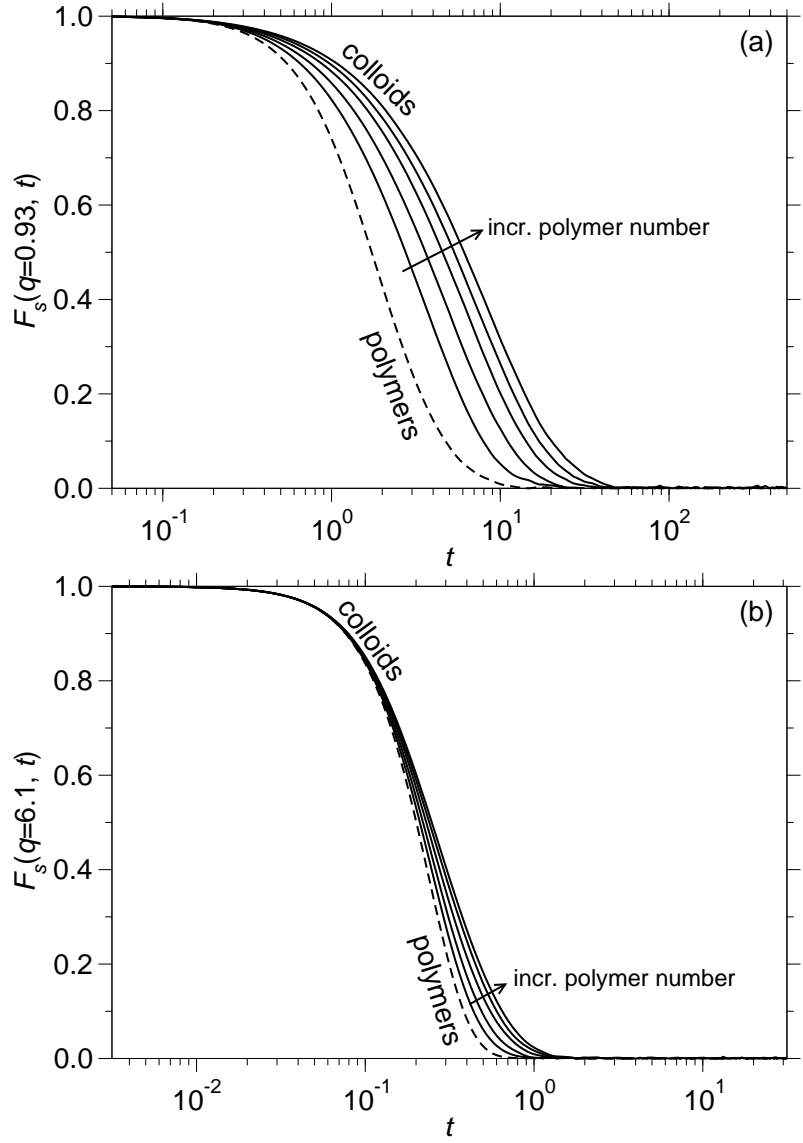

\includegraphics[clip,width=10.5cm]{fig13a}
\includegraphics[clip,width=10.5cm]{fig13b}
\caption{Intermediate incoherent structure factor of colloids and polymers
plotted versus time (note the logarithmic scale) for $\eta_\pp=0.065,
0.129, 0.197, 0.255, 0.318$ and the wave vectors (a) $q=0.93$ and (b)
$q=6.1$. For $F_{\rm s}^\pp (q,t)$ only one curve (dashed) is shown as it
hardly changes with polymer concentration.
\label{fig:10}}
\end{figure}

From the MD runs it is straightforward to obtain the incoherent
intermediate scattering functions $F_s^\alpha (q,t)$ defined as
($\alpha=\rma, \rmb$)
\begin{equation}\label{eq:26}
F_s^\alpha (q,t)=\frac{1}{N_\alpha}
\sum_{i\in\alpha}
\langle\exp(-i\myvec{q}\cdot[\myvec{r}_i(t)-\myvec{r}_i(0)])\rangle
\end{equation}
as well as time-displaced mean square displacements of the particles
\begin{equation}\label{eq:27}
g_\alpha(t)=\frac{1}{N_\alpha}\sum_{i\in\alpha}
\langle[\myvec{r}_{i,\alpha}(t)-\myvec{r}_{i,\alpha}(0)]^2\rangle\,.
\end{equation}
In the MD framework the average $\langle\cdots\rangle$ stands for an
average over the origins of time, $t=0$ (we have used 8 statistically
independent runs and two time origins per run, thus we have averaged over
$16$ time origins).  Figure \ref{fig:10} shows typical data for both
small and large $q$.  For the colloids there is some uniform slowing
down of $F_s^\alpha(q,t)$ at small $q$ as $N_\pp$ increases, while
for large $q$ (near the first peak of $S_{\alpha\beta}(q)$) the decay
occurs in two parts: the first part (for $F_s^\cc(q,t) \gtrsim 0.8$) is
basically independent of $N_\pp$, while for $F_s^\cc(q,t)\lesssim 0.5$
the curves distinctly splay out.  In contrast, the analogous function
for the polymers $F_s^\pp(q,t)$ seems to be practically independent of
$N_\pp$, irrespective of $q$.

\begin{figure}
\includegraphics[clip,width=12.5cm]{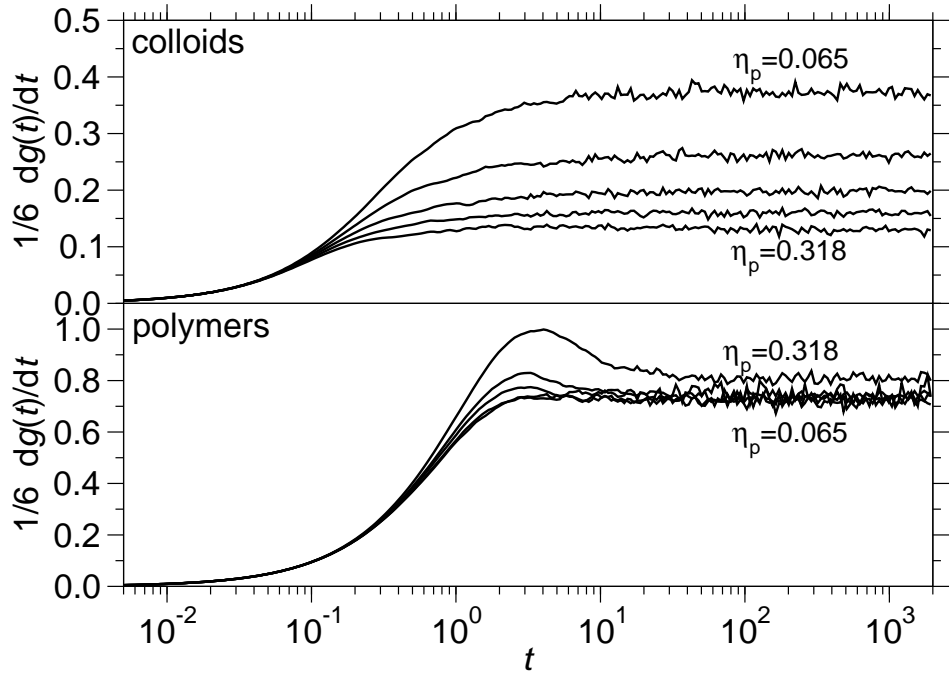}
\caption{Plot of $(1/6)\,\text{d}g_\alpha(t)/\text{d}t$ versus $t$
(note the logarithmic scale) for colloids (upper part) and polymers
(lower part) for the same choices of $\eta_\pp$ as in Fig.~\ref{fig:10}.
\label{fig:11}}
\end{figure}

\begin{figure}
\includegraphics[clip,width=12.5cm]{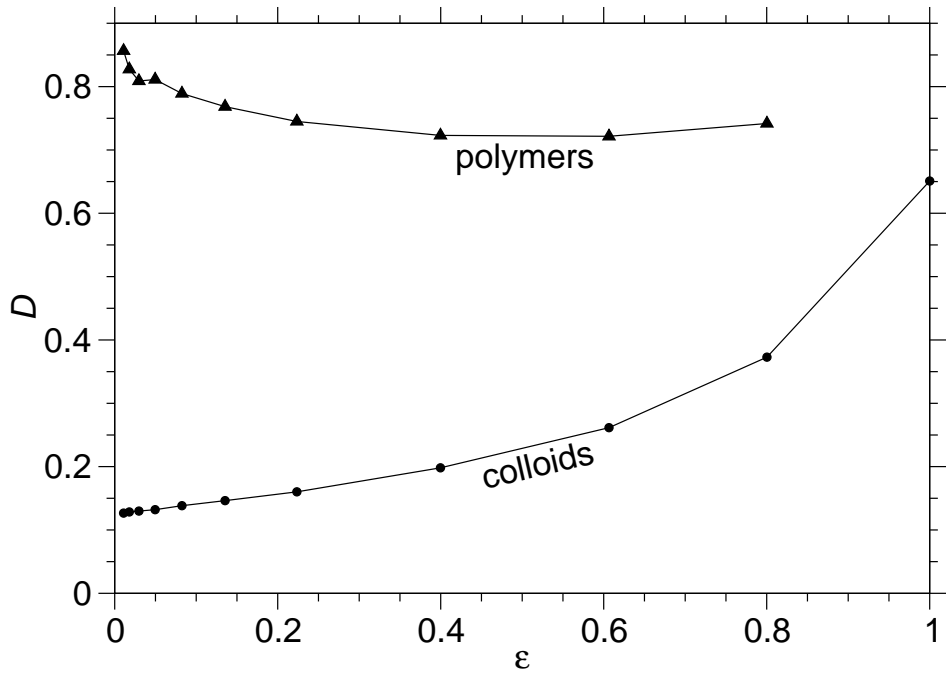}
\caption{Selfdiffusion constants of polymers and colloids
at $\eta_\cc=\eta_{\cc,\crit}=0.150$ plotted versus
$\epsilon=1-\eta_\pp/\eta_{\pp,\crit}$.
\label{fig:12}}
\end{figure}

A similar asymmetry between the dynamics of colloids and polymers is
also seen in the mean square displacements.  Since we expect for large
times the Einstein relation to hold,
\begin{equation}\label{eq:28}
  g_\alpha(t)=6D_\alpha t\,,\quad t\rightarrow\infty\,,
\end{equation}
we analyze the derivative $(1/6)\text{d} g_\alpha(t)/\text{d} t$
(Fig.~\ref{fig:11}).  From the plateau of this quantity at large times,
one can see that $g_\alpha(t)$ approaches its asymptotic behavior for
colloids monotonically while for polymers there is an overshoot for
intermediate times, $1<t<10$.  In the regime of this transient maximum
the data depends rather distinctly on $N_\pp$.  In the asymptotic regime
($t\rightarrow\infty$) the dependence is much weaker.  The time range
where this overshoot occurs is related to the crossover from ballistic
to diffusive motion.  For $t\ll 1$ both colloids and polymers show a
ballistic behavior, $g_\alpha\propto t^2$, as expected \cite{r38,r39}.
Of course, no such behavior is expected for real colloid-polymer mixtures,
where the solvent molecules (no explicit solvent is included in our
simulations, of course) damp out the ``free flight'' motion present in
our model.  Instead, one would find another diffusive motion controlled
by the solvent viscosity.  Figure \ref{fig:12} shows that the resulting
selfdiffusion constants are of similar magnitude for small $N_\pp$ (very
far from $N_{\pp,\crit}$) but differ by almost an order of magnitude
when $N_{\pp,\crit}$ is approached.

Finally, we consider the interdiffusion between colloids and polymers.
Defining the center of mass coordinate of the particles of species
$\alpha$ as $\myvec{R}_\alpha(t)$, we note that interdiffusion is related
to the following mean square displacement ($m_\rma=m_\rmb$) \cite{r41,r64}
\begin{align}
g_\text{int}(t) & =
\left\langle[\myvec{r}_\text{int}(t)-\myvec{r}_\text{int}(0)]^2\right\rangle \nonumber\\
& \equiv \left(1+\frac{N_\rma}{N_\rmb}\right)^2 \frac{N_\rma N_\rmb}{N_\rma+N_\rmb}
\left\langle[\myvec{R}_\rma(t)-\myvec{R}_\rma(0)]^2\right\rangle    \label{eq:29} \,.
\end{align}
Note that $\myvec{R}_\rma (t)-\myvec{R}_\rma (0)$ is computed via the
integral $\int_0^t \myvec{V}_\rma(t')\text{d} t'$ with
$\myvec{V}_\rma(t)=N_\rma^{-1}\sum_i^{N_\rma}\myvec{v}_i(t)$ the
center of mass velocity of component $\rma$ and $\myvec{v}_i(t)$ the
velocity of particle $i$ at time $t$.  In this manner one obtains the
difference $\myvec{R}_\rma(t)-\myvec{R}_\rma(0)$ in an origin independent
representation \cite{r38,r64}.  The Onsager coefficient $\Lambda$
relating to interdiffusion can be expressed as
\begin{equation}\label{eq:30}
\Lambda=\lim_{t\rightarrow\infty} \Lambda(t)\,,\quad
\Lambda(t)=(6t)^{-1}g_\text{int}(t)\,.
\end{equation}
The interdiffusion constant $D_{\rma\rmb}$, which describes how
concentration fluctuations in the binary (A,B)-system relax, is then given
as the ratio of the Onsager coefficient $\Lambda$ and the ``concentration
susceptibility'' $\chi_{CC}$, where
\begin{equation}\label{eq:31}
D_{\rma\rmb}=\frac{x_\rma(1-x_\rma)}{\kb T \chi_{CC}}\Lambda   \;,\quad 
\chi_{CC}=\frac{S_{CC}(q=0)}{\kb T}\,.
\end{equation}
Note that theory \cite{r65a,r65b,r66a,r66b,r67} predicts that $\Lambda$
contains two terms, a background term $\Lambda_b$ which is nonsingular
and stays finite at the critical point and a critical term $\Delta\Lambda$
which diverges at the critical point,
\begin{equation}\label{eq:32}
\Lambda=\Lambda_b+\Delta\Lambda\,,\quad
\Delta\Lambda\propto 
\left(1-\frac{\eta_\pp}{\eta_{\pp,\crit}}\right)^{-\nu_\lambda}
\end{equation}
with an exponent $\nu_\lambda\approx 0.567$ \cite{r68,r69,r70}.  In fact,
a recent MD study of the critical dynamics  of the symmetric binary
Lennard-Jones mixture \cite{r44,r45} yielded results compatible with
this theoretical prediction, Eq.~\eqref{eq:32}.  This allows to estimate
the noncritical background term $\Lambda_b$ at the critical point, too.
Thus, it is also of great interest to study the behavior of $\Lambda$
when we approach the critical point of our model (Fig.~\ref{fig:14}).
Here, we have also included the simple prediction of the Darken equation
\cite{r71},
\begin{equation}\label{eq:33}
\Lambda=x_\rma D_\rmb+(1-x_\rma) D_\rma\,.
\end{equation}
While very far from criticality $(1-\eta_\pp/\eta_{\pp,\crit})\ge
0.6$, Eq.~\eqref{eq:33} indeed describes the simulation results
accurately, it underestimates $\Lambda$ strongly for $\eta_\pp$
closer to $\eta_{\pp,\crit}$, and clearly Eq.~\eqref{eq:33} violates
Eq.~\eqref{eq:32}.  Thus, Darken's equation \cite{r71} fails near the
critical point of a fluid binary mixture as it was already noted for
the binary Lennard-Jones mixture \cite{r72}.

\begin{figure}
\includegraphics[clip,width=15.5cm]{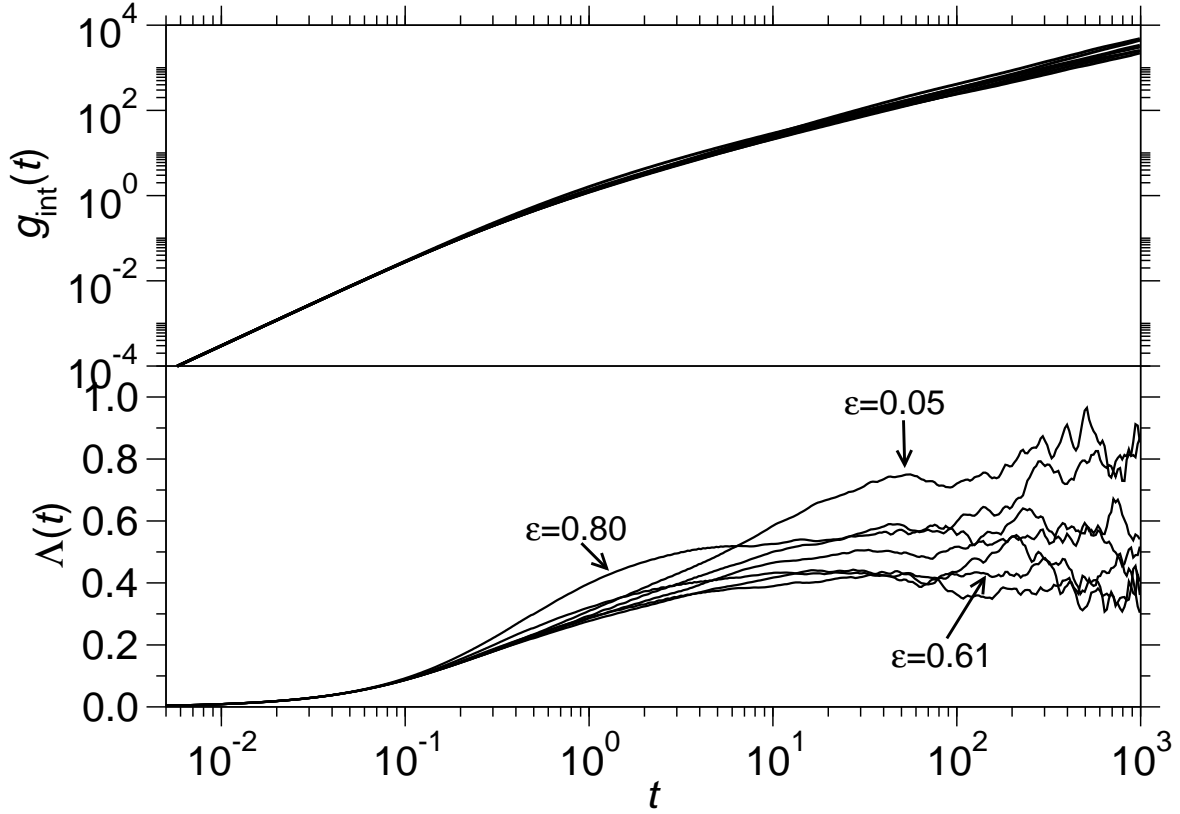}
\caption{Mean square displacement, Eq.~\eqref{eq:29}, relating to
interdiffusion (upper part) and its time derivative $\Lambda(t)$,
Eq.~\eqref{eq:30}, (lower part). The data shown corresponds to the same
values of $\epsilon$ as in Fig.~\ref{fig:9}.
\label{fig:13}}
\end{figure}

\begin{figure}
\includegraphics[clip,width=15.5cm]{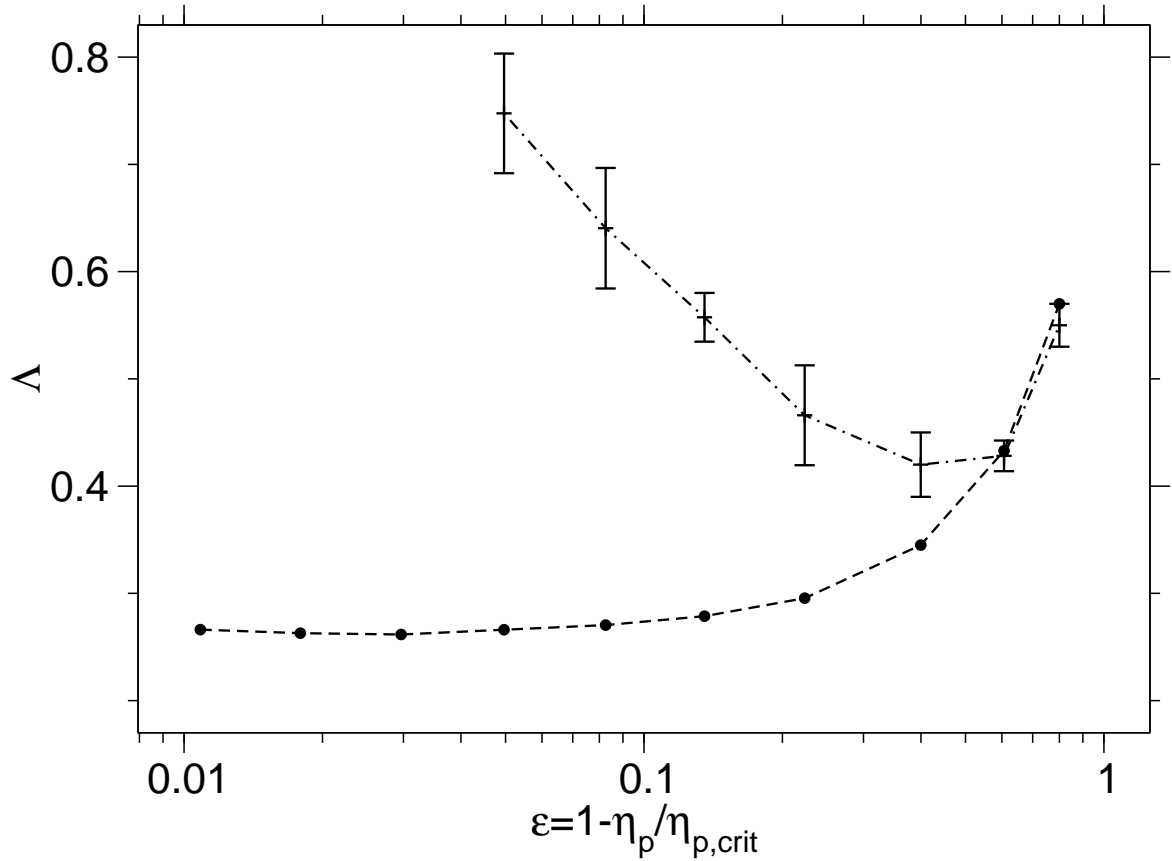}
\caption{Onsager coefficient $\Lambda$ for interdiffusion plotted versus
$\epsilon=1-\eta_\pp/\eta_{\pp,\crit}$ (symbols with error bars). Full
circles show the prediction of the Darken equation \eqref{eq:33}.
\label{fig:14}}
\end{figure}

Thus, we see from Fig.~\ref{fig:14} that for our asymmetric mixture we
also find evidence for a singular behavior of the Onsager coefficient
for interdiffusion.  However, the statistical accuracy of the data
for $\Lambda$ does not warrant an attempt to estimate the dynamic
critical exponent $\nu_\lambda$ (in particular since this is rather
difficult here to estimate $\Lambda_b$).  The statistical effort
invested is just enough to allow an approach of criticality up to about
$\epsilon=1-N_\pp/N_{\pp,\crit}\approx 0.03$, but not closer.  In order
to allow meaningful estimates of $\xi$, $\chi_{CC}$, and $\Lambda$,
the time $\tau_\text{run}$ of a simulation run must be at least about an
order of magnitude longer than the time $\tau$ needed for a concentration
fluctuation to relax via interdiffusion.  This time is
\begin{equation}\label{eq:34}
\tau=(6 D_{\rma\rmb})^{-1} \xi^2  = \frac{\kb T \chi \xi^2}{6 \Lambda}\,.
\end{equation}
From Figs.~\ref{fig:9}(a),(b) and \ref{fig:14} we see for $\epsilon=0.03$
that $\kb T\chi\approx 40$, $\xi\approx 6$, and $\Delta\approx 1$,
yielding $\tau\approx 240$.  Since $\tau_\text{run}=2500$, the run at
$\epsilon=0.03$ is just long enough, but data closer to criticality
cannot be used.  The estimate Eq.~\eqref{eq:34} is compatible with a
direct examination of $\Lambda(t)$, Fig.~\ref{fig:13}, where we see that
far away from criticality a plateau is only reached when $\tau\approx100$.

Another condition for the validity of our result is that the initial
periodicity width $L_\text{init}=9$ has fully relaxed.  This equilibration
time of our system is estimated in analogy to Eq.~\eqref{eq:34} as
$\tau_\text{eq}=(6D_{\rma\rmb})^{-1} L_\text{init}^2 \approx 540$
for $\epsilon=0.03$.  The actual equilibration time of $10^3$ MD time
units indeed exceeds this estimate by a factor of about two.  So our
data should be valid but it is hardly possible to approach criticality
closer.  Finally, since no attempt of a finite size scaling analysis of
the dynamical properties is made here (unlike \cite{r44,r45}), we have to
require that $L\gg 2\xi$ at the states of interest.  Though this condition
holds for $\epsilon=0.03$, it would fail if we approach the critical point
much closer.  From this discussion we see that a substantially larger
computational effort would be required for a more detailed analysis of
the dynamic critical behavior of this model.

\section{Summary}\label{sec:conclusion}
In this paper, a model of colloid-polymer mixtures has been introduced and
studied, which uses continuous potentials between all types of particles.
Nevertheless, it still resembles closely the Asakura-Oosawa (AO)
model, as far as static properties are concerned.  The chosen potentials
(Eqs.~(\ref{eq:ljpotential})-(\ref{eq:6})) are clearly somewhat arbitrary:
the choice of these potentials was motivated by the desire that the
model should be suitable for grand-canonical Monte Carlo methods to
accurately establish the phase diagram (Figs.~\ref{fig:4}, \ref{fig:5}).
In addition, it should allow for a convenient and physically meaningful
application of Molecular Dynamics techniques.  In this way, both static
and dynamic behavior of such a phase-separating strongly asymmetric
binary mixture, where phase separation is mainly driven by entropic
depletion effects, has become accessible to a computer simulation study.

Previous simulation studies have mostly been concerned with the
phase diagram of the AO model and related models as well as the
interfacial tension between coexisting polymer-rich and colloid-rich
phases.  The present work contains a detailed analysis of the various
static structure factors $S_{\rma\rma} (q)$, $S_{\rmb\rmb}(q)$ and
$S_{\rma\rmb}(q)$ that one can define in such a binary (AB) mixture,
and suitable linear combinations of them that single out the order
parameter of the unmixing transition.  Unlike other models of (almost
incompressible) binary mixtures, in the present system the relative
concentration of one species is not the proper order parameter.  Instead
the order parameter is a nontrivial combination of polymer density and
colloid density fluctuations, which can be found from diagonalizing the
structure factor matrix.  From this analysis, we can study the onset of
the critical divergence of both the order parameter ``susceptibility''
and correlation length.  Roughly, these results are compatible with the
expected Ising-like criticality.  Fine details such as whether Fisher
renormalization of critical exponents occur can, unfortunately, not be
clarified, since our data are restricted to relative distances from the
critical point exceeding $0.04$.  Of course, for reliable statements
on critical exponents data somewhat closer to the critical point are
indispensable, but at present not yet available.

A central part of our study concerns the analysis of time-dependent
quantities, intermediate incoherent structure factors and mean square
displacements, and their analysis.  While the self-diffusion constant of
the colloids is decreasing monotonously with increasing polymer density,
surprisingly the self-diffusion constant of the polymers shows a slight
increase.  Thus, there is a pronounced dynamic asymmetry of our model.
The Onsager coefficient relating to interdiffusion is also obtained, and
qualitative evidence for a critical divergence is found, thus invalidating
the simple Darken equation for this system.  However, the present data
do not yet allow an accurate estimation of dynamic critical exponents for
the colloid-polymer mixture.  More efficient algorithms (or significantly
faster computers) will be needed for a more definite study of critical
behavior in our model system.  Nevertheless, we hope that our study will
motivate related experimental work on the dynamics of colloid-polymer
mixtures, to which some of our findings could be compared directly.

\section*{Acknowledgments}
This work received financial support from the Deutsche
Forschungsgemeinschaft, TR6/A5. Computing time on the JUMP at the NIC
J\"ulich is gratefully acknowledged.

\newpage

\bibliography{paper_zausch_jcp}

\printfigures

\end{document}